\documentclass[twocolumn]{aastex631}

\accepted{\today}

\begin{document}

\title{Tracing Pebble Drift History in Two Protoplanetary Disks with CO Enhancement}

\correspondingauthor{Tayt Armitage}
\email{tarmitage@wisc.edu}

\author[0000-0003-3190-8890]{Tayt Armitage}
\affiliation{Department of Astronomy, University of Wisconsin-Madison, 
475 N Charter St, Madison, WI 53706, USA}

\author[0009-0008-8176-1974]{Joe Williams}
\affiliation{School of Physics and Astronomy, University of Exeter, Stocker Road, Exeter, EX4 4QL, UK}

\author[0000-0002-0661-7517]{Ke Zhang}
\affiliation{Department of Astronomy, University of Wisconsin-Madison, 
475 N Charter St, Madison, WI 53706, USA}

\author[0000-0002-3291-6887]{Sebastiaan Krijt}
\affiliation{School of Physics and Astronomy, University of Exeter, Stocker Road, Exeter, EX4 4QL, UK}

\author[0000-0002-8623-9703]{Leon Trapman}
\affiliation{Department of Astronomy, University of Wisconsin-Madison, 
475 N Charter St, Madison, WI 53706, USA}

\author[0000-0002-0364-937X]{Richard A. Booth}
\affiliation{School of Physics and Astronomy, University of Leeds, Leeds, LS2 9JT, UK}

\author[0000-0003-1534-5186]{Richard Teague}
\affiliation{Department of Earth, Atmospheric, and Planetary Sciences, Massachusetts Institute of Technology, Cambridge, MA 02139, USA}

\author[0000-0003-1413-1776]{Charles J. Law}
\altaffiliation{NASA Hubble Fellowship Sagan Fellow}
\affiliation{Department of Astronomy, University of Virginia, Charlottesville, VA 22904, USA}

\author[0000-0001-8642-1786]{Chunhua Qi}
\affiliation{Institute for Astrophysical Research, Boston University, 725 Commonwealth Avenue, Boston, MA 02215, USA}

\author[0000-0003-1526-7587]{David J. Wilner}
\affiliation{Center for Astrophysics, Harvard \& Smithsonian, 60 Garden St., Cambridge, MA 02138, USA}

\author[0000-0001-8798-1347]{Karin I. \"Oberg}
\affiliation{Center for Astrophysics, Harvard \& Smithsonian, 60 Garden St., Cambridge, MA 02138, USA}

\author[0000-0003-4179-6394]{Edwin A. Bergin}
\affiliation{Department of Astronomy, University of Michigan, 1085 S. University Avenue, Ann Arbor, MI 48109, USA}

\author[0000-0003-2253-2270]{Sean M. Andrews}
\affiliation{Center for Astrophysics, Harvard \& Smithsonian, 60 Garden St., Cambridge, MA 02138, USA}

\author[0000-0003-1837-3772]{Romane Le Gal}
\affiliation{Institut de Planétologie et d'Astrophysique de Grenoble, Institut de Radioastronomie Millimétrique,Université Grenoble Alpes, Saint-Martin-d'Hères, Auvergne Rhône-Alpes, FR}

\author[0000-0002-7607-719X]{Feng Long}
\affiliation{Lunar and Planetary Laboratory, University of Arizona, Tucson, AZ 85721, USA}

\author[0000-0001-6947-6072]{Jane Huang}
\affiliation{Department of Astronomy, Columbia University, 538 West 120th Street, Pupin Hall, New York, NY 10027, USA}

\author[0000-0001-7258-770X]{Jaehan Bae}
\affiliation{Department of Astronomy, University of Florida, Gainesville, FL, US}

\author[0000-0002-2692-7862]{Felipe Alarcón}
\affiliation{Dipartimento di Fisica, Università degli Studi di Milano, Via Celoria 16, 20133 Milano, Italy}



\begin{abstract}
Pebble drift is an important mechanism for supplying the materials needed to build planets in the inner region of protoplanetary disks. Thus, constraining pebble drift's timescales and mass flux is essential to understanding planet formation history. Current pebble drift models suggest pebble fluxes can be constrained from the enhancement of gaseous volatile abundances when icy pebbles sublimate after drifting across key snowlines. In this work, we present ALMA observations of spatially resolved $^{13}$C$^{18}$O $J$=2-1 line emission inside the midplane CO snowline of the HD 163296 and MWC 480 protoplanetary disks. We use radiative transfer and thermochemical models to constrain the spatial distribution of CO gas column density. We find that both disks display centrally peaked CO abundance enhancement of up to ten times of ISM abundance levels. For HD 163296 and MWC 480, the inferred enhancements require 250-350 M$_{\oplus}$ and 480-660 M$_{\oplus}$ of pebbles to have drifted across their CO snowlines, respectively.
These ranges fall within cumulative pebble mass flux ranges to grow gas giants in the interior to the CO snowline. The centrally peaked CO enhancement is unexpected in current pebble drift models, which predict CO enhancement peaks at the CO snowline or is uniform inside the snowline.  We propose two hypotheses to explain the centrally-peaked CO enhancement, including a large CO desorption distance and CO trapped in water ice.  By testing both hypotheses with the 1D gas and dust evolution code \texttt{chemcomp}, we find that volatile trapping (about 30\%) best reproduces the centrally peaked CO enhancement observed. 
\end{abstract}


\section{Introduction} \label{sec:intro}
The pebble accretion model of planet formation has become increasingly favored as it enables fast formation of gas giants within the gas disk lifetime \citep[e.g.,][]{2015A&A...582A.112B}. This model states that planetary embryos experience rapid growth within the first few Myr due to the accretion of millimeter to centimeter-sized pebbles within a gas-rich environment \citep{Lambrechts_2019}. In a static-disk scenario (i.e., no transport of disk material), pebble accretion can explain the formation of gas giants in the outer regions of a disk, but due to insufficient dust budgets, they cannot form in regions closer to their host star \citep{2015A&A...582A.112B}. In order to drive the formation of giant planets in the inner regions of a protoplanetary disk additional solid material is required to be brought into the inner disk. This transport mechanism is known as pebble drift, which occurs when dust grains in the outer disk grow into ice-coated pebbles and settle into the midplane. The gas within the disk then imparts a drag force on the pebbles, causing them to drift radially inward, which then supplies the required material to drive rapid planet growth within the disk lifetime \citep{Whipple_1972,Cielsa&Cuzzi_2006}. 

Recent simulations have shown that pebble drift is critical in determining the populations of planets that form within a disk \citep{Ormel_2017,Lambrechts_2019,Johansen_2021}. In these works, over a hundred Earth masses of pebbles are required to drift into the inner disk within the first few Myr to form super-Earths. Observationally, the total pebble mass flux and efficiency of pebble drift remain largely unconstrained. Due to high dust opacity and uncertainty in pebble size, pebble mass flux cannot be directly constrained from dust continuum images \citep{Birnstiel_2018,Liu_2022}. 

However, this does not mean pebble drift cannot be observed. One option is that gas tracers from sublimating ices can be used to infer a cumulative pebble flux. As pebbles are transported throughout the midplane, they cross thresholds within the disk known as snowlines, where volatile ices (H$_2$O, CO, CO$_2$) on the pebble mantle sublimate and subsequently enrich the gas phase volatile abundances around their respective snowlines \citep[e.g.,][]{Cuzzi&Zahnle_2004,Oberg&Bergin_2016,Booth_2017,Booth&Ilee_2019}. CO has the potential to be a particularly useful tracer for pebble drift.  Simulations show that the sublimation of CO can create a local peak by a factor of a few in the CO gas column density in the region around the CO snowline \citep[e.g.,][]{Stammler_2017,Krijt_2018, Krijt_2020}. This CO abundance enhancement, if isolated, can serve as a ``smoking gun" for pebble drift. Depending on the timescales of the pebble drift, the distribution of the volatile enhancement can then be used to determine the pebble drift history of each disk. Work done by \citet{Cuzzi&Zahnle_2004} introduces this effect where under long timescales, uniform enhancement of volatiles can occur interior to snowlines, or local plumes can indicate active volatile delivery. Alongside local plumes at the initial snowline, later enhancement can occur if volatiles are trapped within ices with higher sublimation temperatures \citep{Bar-nun_1985}. Recent work has shown that in the case for CO alone, it can be trapped in water and CO$_{2}$ ice, leading to delayed sublimation and CO enhancement further in the disk, which may be critical for shaping the C/O ratios in planet-forming regions \citep{Simon_2019,Ligterink_2024}.

Significant CO enhancements have been observed interior to the CO snowlines of two well-characterized protoplanetary disks around HD 163296 and MWC 480 through C$^{17}$O and C$^{18}$O line emission \citep{Booth_2019_HD163,Zhang_2020,Loomis_2020, Zhang_2021}. It is hypothesized that the CO enhancement is due to large-scale pebble drift \citep{Zhang_2020, Zhang_2021}. However, previous observations lacked the resolution to adequately resolve the midplane CO enhancement interior to the CO snowline. Thus, a detailed assessment of the pebble drift history could not be performed in these initial studies.

In this work, we present the first radially resolved $^{13}$C$^{18}$O observations interior to the CO-snowlines of the HD 163296 and MWC 480 disks.  As $^{13}$C$^{18}$O is optically thin, it allows us to analyze the midplane CO enhancement and to provide new constraints on the pebble drift history of both disks. In \S\ref{sec:obs}, we describe the new spatially resolved $^{13}$C$^{18}$O ALMA observations. In \S\ref{sec:methods}, we explain the methodology used to model the CO column density distribution throughout each disk. The results of these models are shown in \S\ref{sec:results}. We then estimate the cumulative pebble flux for each disk and discuss possible mechanisms to produce the observed enhancement profile in each disk in \S\ref{sec:discussion}. We summarize our findings and present future directions in \S\ref{sec:conclusion}.

\section{Observations} \label{sec:obs}

The observations were carried out with the ALMA 12-meter array between 05 July 2022 and 07 August 2022 (project ID: 2021.1.00899.S, PI. K. Zhang). The total on-source integration time on HD 163296 and MWC 480 were 107 and 158.4 minutes, respectively. Baselines ranged from 15 to 2617\,m. The spectral setup consists of four spectral line windows centered at 195.948, 195.141, 209.224, 209.413\,GHz, respectively, covering CS\,(4-3), CH$_3$OH (4$_{1,3}$-3$_{1,2}$), HC$_3$N (23-22), and $^{13}$C$^{18}$O (2-1) with a velocity resolution of 0.2\,km s$^{-1}$. The key science line was $^{13}$C$^{18}$O (2-1). Two continuum spectral windows with 1875\,MHz bandwidth each are used to help improve calibrations. Quasar J1742-1517 and J0438+3004 were used as phase calibrators for HD 163296 and MWC 480, respectively. 

All execution blocks (EBs) are initially calibrated using the Common Astronomy Software Application (CASA; \citealt{CASA2022}) pipeline version 6.2.1.7. After the standard pipeline calibration, we conducted self-calibration using CASA version 6.4 to improve the signal-to-noise (SNR) of the data. All EBs are re-aligned to a common center with \texttt{fixplanets} and \texttt{fixvis} tasks. Three rounds of phase and one round of amplitude self-calibration were performed, using \texttt{solint=`inf',`120s',`60s',`inf'}, respectively. The final continnuum images's peak SNR improved a factor of four compared ones before self-cal.

After applying the self-calibration, we then use the task \texttt{uvcontsub} to subtract the continuum with all spectral windows together, providing a continuum-subtracted measurement set for each source.

We utilized the \texttt{tclean} task in CASA with the multiscale clean option to generate line images. We applied tapering to visibilities to generate a circular beam that is useful to compare radial intensity profiles of molecular line emission. With a fixed robust parameter of 0.3, we calculate the tapering profiles needed to deliver a circularized CLEAN beam, and applied the tapering profiles to \texttt{tclean} with the \texttt{uvtaper} keyword. The resulting beam sizes are 0\farcs3 for HD 163296 and 0\farcs4 for MWC 480 line images.

\begin{figure*}
    \centering
    \includegraphics[width = \textwidth
]{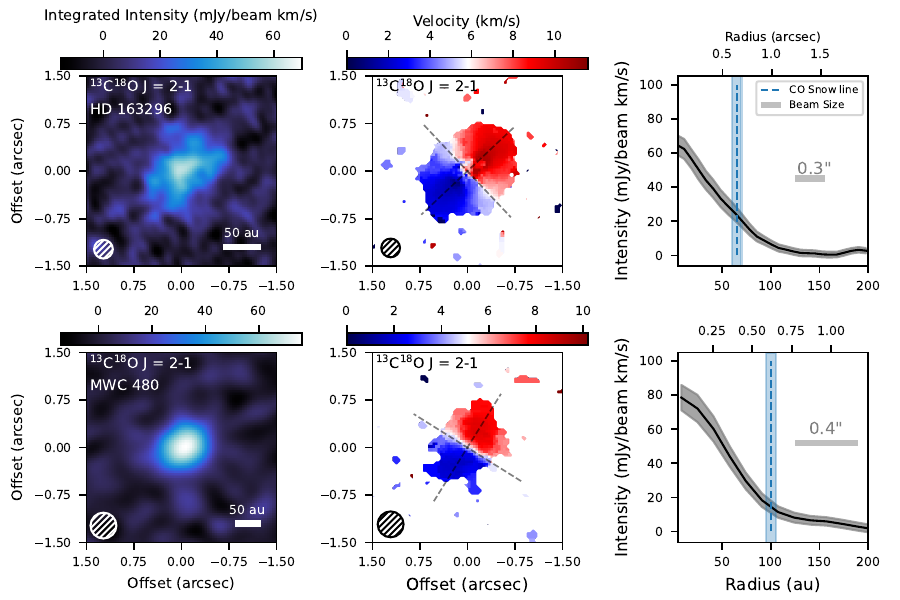}
    \caption{This figure details $^{13}$C$^{18}$O observations of both the HD 163296 (Top) and MWC 480 protoplanetary disks (bottom). The first column depicts the moment zero map, the middle shows the moment one map, and the radial profile of the integrated spectra in the final column. In the center column we include dashed lines to indicate the major and minor axes of each disk. Shown in blue are the midplane CO snowlines at 60 au for the HD 163296 disk and 100 au for the MWC 480 disk. These values were derived by \citet{Zhang_2021}. }
    \label{fig:observations}
\end{figure*}

We utilize \texttt{bettermoments} \citep{2018RNAAS...2c.173T} and \texttt{GoFish} \citep{Teague2019} packages to generate moment 0 maps, moment 1 maps, and radial integrated intensity profiles for each disk. The results are shown in Figure \ref{fig:observations}. We report a peak intensity of 20.4 mJy/beam and noise level of 3.1 mJy/beam for the $^{13}$C$^{18}$O channel maps of the HD 163296 disk, and a peak intensity of 19.9 mJy/beam and a noise level of 1.9 mJy/beam for the MWC 480 disk.

Additionally, we use the observations of multiple CO isotopologues taken with the ALMA-MAPS program \citep{Oberg_2021} for HD 163296 and MWC 480, including C$^{18}$O J = 2-1 transition at 0\farcs15 resolution and C$^{17}$O J = 1-0 transition at 0\farcs3 resolution. Details of these observations and calibrations can be found in \citet{Zhang_2021, Law_2021, Czekala_2021}.

\section{Modeling Methods} \label{sec:methods}

In this work, we adopt the modeling procedures introduced in \citet{Zhang_2021}. In the following sections, we briefly reiterate the core modeling principles that remain the same and direct the reader to \citet{Zhang_2021} for more in-depth explanations of certain modeling set-ups. Along with modeling procedures, we also adopt the same initial chemical network, opacity structure, and stellar parameters. The stellar and disk parameters used for our models can be found in Table \ref{table:modelparams}. We then make slight modifications to the density structure of each disk in Section \ref{sec:duststructure}.

\subsection{Modeling Dust Disk Structure}{\label{sec:duststructure}}
Following the workflow of \citet{Zhang_2021}, we begin our analysis by developing dust structure models for both disks. We begin by using the ray-tracing module of the radiative transfer software \texttt{RADMC3D} \citep{2012ascl.soft02015D} to generate initial model millimeter continuum images using the stellar spectrum, dust opacity, and best-fit dust surface density profiles presented in \citet{Zhang_2021}. The dust considered in our model is then broken into two components: a large and small dust grain population. The disk inclinations and position angles (PA) are shown in Table \ref{table:inputs}. We then deviate from the previous work by incorporating
anisotropic photon scattering into our models. We place an emphasis on this as we aim to characterize the CO enhancement interior to 60 au. In this region, the dust emission is expected to become less optically thin \citep{Bosman_2021_innerdisk,Sierra_2021}, and therefore, scattering effects become important in properly characterizing the dust population \citep{Zhu_2019}

The anisotropic scattering is governed by the Heyney-Greenstein Scattering function \citep{Henyey_1941}, which is given by 

\begin{equation}
    \phi(\mu) = \frac{1- g^{2}}{(1 + g^{2} - 2g\mu)^{\frac{3}{2}}}
\end{equation}
where g is determined by our input dust size distribution and $\mu$ is the cosine of the scattering angle. This scattering mode was chosen as it presents a more realistic model of the dust structure while maintaining reasonable computation time. It is also important to mention that in modeling the continuum image of the HD 163296 disk we increase the photon counts for the radiative transfer modeling to overcome the high opacity presented in the inner 10 au.

Once completed, the model image is then convolved with a Gaussian to reflect the beam size and position angle in order to reproduce the observed 1.3\,mm continuum. Using the radial\_profile module of \texttt{GoFish} we then generate radial intensity profiles for the model image cube, which are used to compare to those presented in the DSHARP program \citet{Andrews_2018_dsharp} and \citet{2018ApJ...869...17L}. This comparison then aids in fine-tuning the large dust grain surface density for both disks, where the percent error between the model and observed radial intensity serves as a local scaling factor. The scaling factor is then applied to the large dust grain surface density profile. The updated surface density profiles are used as input parameters for \texttt{RADMC3D} where we then calculate a new dust temperature structure. Finally, we return to the ray-tracing module to generate new model continuum images and radial profiles. This process is iterated until our models sufficiently reproduce observations. We present our final model continuum images and best-fit dust surface density profiles in Appendix \ref{appendix:A}. We note that as a result of this updated large dust population the masses we report in Table \ref{table:modelparams} are slightly larger than those found in \citet{Zhang_2021}. Our total large dust masses for each disk are  2.4$\times10^{-3}$\,M$_{\odot}$ and 1.52$\times10^{-3}$\, M$_{\odot}$ for HD 163296 and MWC 480 respectively, compared to the previous measurements of 2.31$\times10^{-3}$\,M$_{\odot}$ and 1.42$\times10^{-3}$\,M$_{\odot}$.

\subsection{Gas Surface Density Profile}\label{gas_surface_density}

In this section we briefly go over the methods used to derive the gas surface density profile, which we have adopted from \citet{Zhang_2021}. The mass surface distrubition was set as a self-similar viscous disk \citep{Lynden-Bell&Pringle_1974}, described by the following equation:

\begin{equation}\label{selfsimilar}
    \Sigma \left(R\right) = \Sigma_{c}\left(\frac{R}{R_c}\right)^{-\gamma}\exp\left[-\left(\frac{R}{R_c}\right)^{2-\gamma}\right]
\end{equation}

For our calculations, we adopt the characteristic radius R$_{c}$, gas surface density exponent $\gamma$, and surface density at R$_c$, $\Sigma_c$, from Table 2 of \citet{Zhang_2021}. The values for R$_c$, $\gamma$, and $\Sigma_c$ were each derived using best-fit thermochemical models of optically thick CO (2-1) and $^{13}$CO (1-0) emission (see sec. 3.1.6 of \citet{Zhang_2021}). The vertical structure of the gas was then determined by

\begin{equation}
    \rho_{i}(R,Z) = f_i \frac{\Sigma(R)}{\sqrt{2\pi}H_i(R)} exp[-\frac{1}{2}(\frac{Z}{H_i(R)})^2]
\end{equation}

\begin{equation}
    H_i(R) = \chi_i H_{100}(R/100 au)^\psi
\end{equation}

    Where the scale height H and $\psi$, which characterizes the radial dependence of the scale height were determined by best-fitting models to the mid and far infrared SED. The models of the SED, along with the values for H and $\psi$ are reported in \citet{Zhang_2021}. It is also important to note that as the parameters used to determine the gas surface density profile were derived from CO lines, it introduces a degeneracy between the CO abundance and $\Sigma_{gas}$. To overcome this degeneracy in our analysis we also consider independently derived kinematic gas masses \citep{Trapman_dynamic_disk_mass_2025} and additional molecular observations of each disk \citep{Oberg_2021,Bosman_2021,Law_2021}. Further discussion on this degeneracy is left to section \ref{Robustness}.

Along with the gas surface density we adopt total disk gas mass to for our models determined by \citet{Zhang_2021}, but we note that our global and local gas-to-dust ratios differ. This is due to our additional modeling of the large dust grain population in each of our disks. As such we have global gas-to-dust ratios of 53.8 and 94.7 for the HD 163296 and MWC 480 disks respectively. The local gas-to-dust ratio changes radially depending on the gas and dust density profiles.

\subsection{CO Abundance Modeling} \label{Abundance Modeling}

Once we calculate the dust temperature structures and have our best-fit dust surface density profile, we then compute the CO abundance structure. To do this, we adopt temperature structures computed using the thermochemical software 
\texttt{RAC2D} \citep{Du&Bergin2014}. For a given gas and dust density structure \texttt{RAC2D} self-consistently calculates the gas temperature structure and a time-dependent chemical abundance structure within the disk. For these models we use the initial chemical abundances outlined in Table 1 of \citet{Du&Bergin2014}. In these models we neglect CO isotopologue fractionation because both disks are massive, and thus fractionation is expected to be insignificant \citep{Miotello_2016}. The elemental abundances assumed during our models do not mimic C and O depletion, instead, the resulting CO abundance outputs are scaled using local ISM isotopologue abundance ratios of CO/C$^{18}$O = 570, CO/$^{13}$CO = 69, CO/C$^{17}$O = 2052, and CO/$^{13}$C$^{18}$O = 39330 \citep{Wilson_1999}.

The initial gas density and temperature structures serve as inputs to our  \texttt{RADMC3D} models, where we use the ray-tracing module to generate model-line image cubes. We do this for  $^{13}$C$^{18}$O J = 2-1, C$^{18}$O J = 2-1, and C$^{17}$O J = 1-0. While generating the line image cubes, we incorporate anisotropic photon scattering. Once generated, the model channel maps are binned to match the spectral resolution of our observations. They are then convolved with a Gaussian that has the same beam size as our CO line observations—following the same procedure used by \citet{Zhang_2021}. Once convolved, the \texttt{GoFish} python package is used to produce azimuthally averaged radial intensity profiles of each model image.

We compare our model radial intensity profiles to those generated from our CO ALMA observations. This helps us develop a CO enhancement factor, which is given as the ratio of our computed CO abundance to the standard ISM CO abundance of CO/H$_2$ = 10$^{-4}$. 

In our modeling, we started by using the CO enhancement profiles presented in \citet{Zhang_2021} as inputs to generate simulated line image cubes to compare with observations. Similar to the scaling of the large dust population, the percent error between our model radial intensity profile and the observations at discrete radii is used as a scale factor to vary the CO enhancement factor before generating new model images. To prevent over-fitting, we apply a linear-smoothing function based on the angular resolution of our observations to our CO enhancement profiles with the \texttt{SciPY} \citep{Scipy}. 

We apply the updated CO enhancement factor to revise the CO abundance structure used for our models and generate a new model radial intensity profile. The updated model radial profile is then compared with the observations as was done before. We repeat the above process until our model radial profile falls within the error estimates of our observations. The CO enhancement profiles used to create our best-fit models are adopted to calculate the excess CO within the CO snowline. We implement this procedure independently for each CO line observation for both disks. We discuss the best-fit CO enhancement factors and model radial intensity profiles in section ~\ref{sec:results}.

\begin{deluxetable*}{c c c c c c c c c c c c c }
\tabletypesize{\footnotesize}
\tablecolumns{10}
\tablewidth{0pt}
\tablecaption{ Stellar and Disk Parameters \label{table:modelparams}}
\tablehead{
\colhead{Source} & \colhead{d} &  \colhead{Incl.} & \colhead{PA} & \colhead{T$_{eff}$}
& \colhead{L$_{*}$} & \colhead{(M$_*$)} & \colhead{v$_{sys}$} & \colhead{$\gamma$} & \colhead{R$_{c}$} & \colhead{M$_{mm}$} & \colhead{M$_{\mu m}$} & \colhead{M$_{gas}$} \\  & \colhead{(pc)} & \colhead{(deg)} & \colhead{(deg)} & \colhead{(K)} & \colhead{(L$_{\odot}$)} & \colhead{(M$_{\odot}$)} & \colhead{(km/s)} & & \colhead{(au)} & \colhead{(M$_{\odot}$)} & \colhead{(M$_{\odot}$)} & \colhead{M$_{\odot}$} \\
\colhead{(1)} & \colhead{(2)} & \colhead{(3)} & \colhead{(4)} & \colhead{(5)} & \colhead{(6)} & \colhead{(7)} & \colhead{(8)} & \colhead{(9)} & \colhead{(10)} & \colhead{(11)} & \colhead{(12)}  & \colhead{(13)}}
\startdata
HD 163296 & 101 & 46.7 & 133.3 & 9332 & 17.0 & 2.0 & 5.8 & 0.8 & 165 & 2.40e-3 & 2.00e-4 & 0.14 \\ 
MWC 480 & 162 & 37.0 & 148.0 & 8460 & 21.9 & 2.11 & 5.1 & 1.0 & 200 & 1.52e-3 & 1.69 e-4  & 0.16 \\
\enddata

\tablecomments{Above is a list of stellar, disk, and dust parameters used for our models. Column (1): depicts the sources. Column (2): the distance to the disk. Column (3): is the inclination. Column (4) is the Position Angle. Column (5): is the effective temperature. Column (6): is the luminosity. Column (7) is the stellar mass. Column (8) is the systematic velocity. Column (9) is the surface density exponent. Column (10) is the characteristic radius. Column (11) is the large dust mass budget. Column (12) is the small dust mass budget. Column (13) is the total disk gas mass. The values from Columns 1-10 and 12-13 are sourced from \citet{Zhang_2021}, while Column 10 uses updated values for this paper.}
\label{table:inputs}
\end{deluxetable*}

\section{Results} \label{sec:results}
\begin{figure*}
   \centering
   \includegraphics[width = \textwidth]{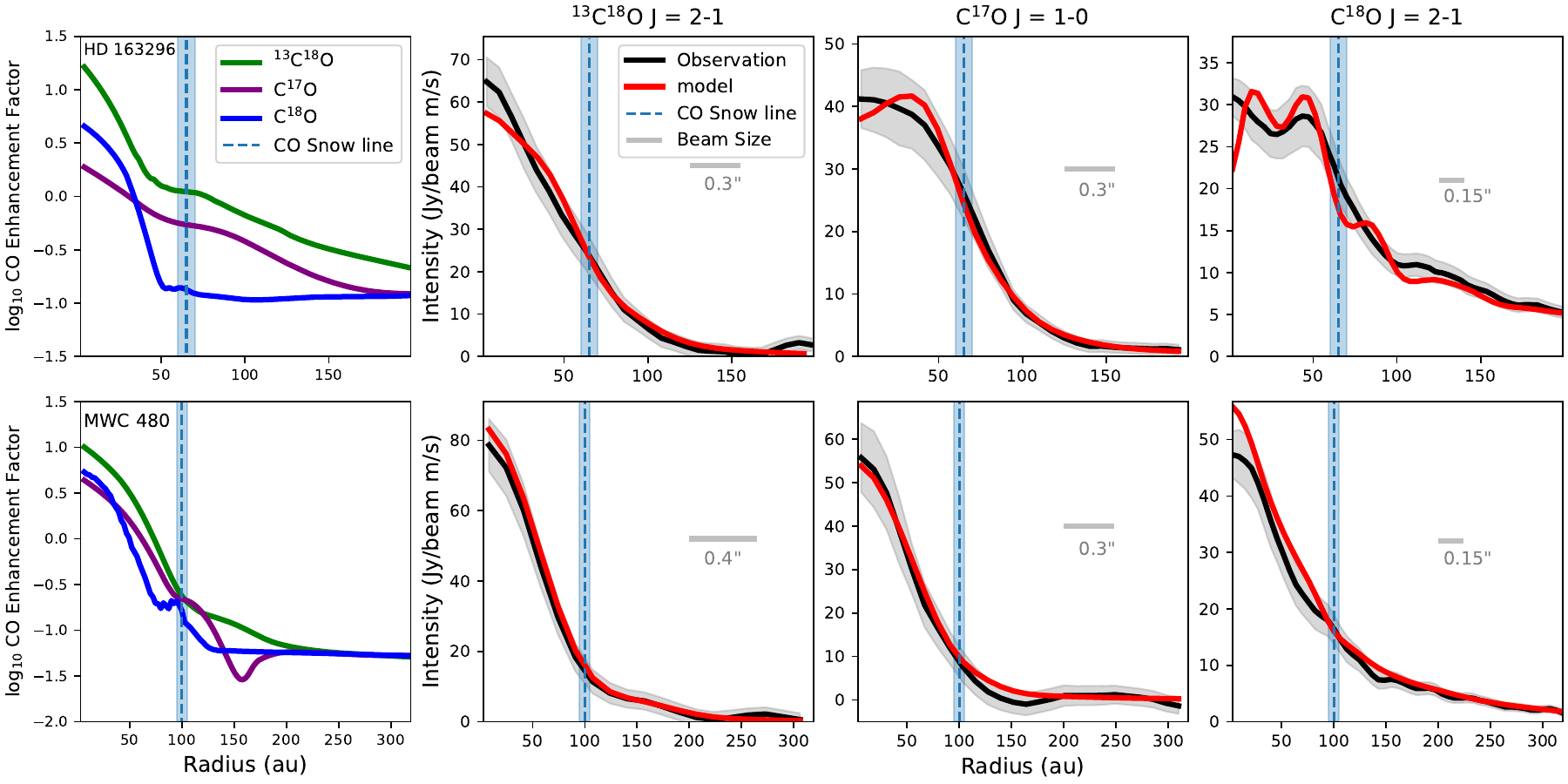}
    \caption{The leftmost panel displays the CO enhancement profile on a logarithmic scale as a function of radius for the three CO isotopologues analyzed for each disk. Each Enhancement profile has been smoothed with the corresponding beam size. The right three panels then display the observed radial intensity profile (black) and our best-fit models (red). We also include the beam size for each observation, along with the location of the midplane CO snowline (light blue).}
    \label{fig:best_models}
\end{figure*}

We present the results of our best-fitting models in Figure \ref{fig:best_models}. As can be seen across each radial profile, every isotopologue of CO displays significant centrally peaked enhancement interior to the snowline. This effect becomes greater as the lines become more optically thin and probe deeper into the midplane.

Looking to the left side of Figure 
\ref{fig:best_models} we showcase the CO enhancement factor in log scale. Each isotopologue requires significant enhancement, with $^{13}$C$^{18}$O being enhanced up to 10x the ISM abundance in the innermost regions of the disk. This effect does differ among observations of C$^{18}$O, C$^{17}$O, and $^{13}$C$^{18}$O. 
The enhancement difference among different isotopologues has larger scattering in the HD 163296 disk than that of the MWC 480 disk. We attribute the difference in enhancement among isotopologues to each isotopologue probing different vertical layers of the disk. This is due to varying optical depth, where $^{13}$C$^{18}$O is the most optically thin line that probes into the deepest layers close to the midplane. This is further predicted by the vertical temperature structure used in our models constrained by the emitting layers of CO isotopologues \citep{Law_2021_temperature}. The $^{13}$C$^{18}$O requires a larger enhancement suggest that the deep layer close to the midplane has a higher CO abundances than the atmosphere layers probed by C$^{18}$O.

In our modeling, it is important to note that for the C$^{18}$O model for the HD 163296 protoplanetary disk deviates from the observed profiles for the innermost 10 au. We believe this is due to the high opacity gradient in the inner few au for C$^{18}$O, which \texttt{RADMC3D} struggles to reproduce. An indicator towards this is that as increasing the photon count in our Monte Carlo simulations reduced noise and improved our model's ability to reproduce the observations. However, extensive computation time is required to simulate enough photons to fully overcome the difficulty in modeling the steep opacity gradient in the early cells such that the integrated intensity radial profile falls within the error bars of the observations. We determine that our best fit is sufficient for this work as this region is already difficult to resolve in the observed data, and thus, it should not affect our analysis.

In addition to the radial integrated intensity profiles presented in Figure \ref{fig:best_models}, we also include modeled integrated spectrum for both disks using \texttt{GoFish}. We present these spectra in Appendix \ref{appendix:A}.

 \subsection{Result Robustness}
\label{Robustness}

In this work we use the $^{13}$C$^{18}$O as a unique tracer of the overall CO enhancement within the protoplanetary disks HD 163296 and MWC 480. It is important to note that the CO enhancement is in reference to the ISM CO/H$_{2}$ abundance ratio, where in each of these disks the H$_{2}$ column density remains difficult to constrain in the inner disk. However in this section we present evidence that the enhancement in the $^{13}$C$^{18}$O is attributed to a uniquely high density column of CO, rather than an overall elevated gas column in the inner disk.

Our constraints on the CO/H$_{2}$ abundances depend on the total disk mass. Our models adopt gas masses of 0.14 M$_{\odot}$ and 0.16 M$_{\odot}$ for HD 163296 and MWC 480 respectively based upon the work of \citet{Zhang_2021}. We have compared these values to kinematically derived disk masses, and found them to be consistent with literature.
 Work done by \citet{Trapman_dynamic_disk_mass_2025} found that gas disk masses for HD 163296 and MWC 480 derived via line flux and kinematics were consistent to within 1-2 $\sigma$, placing finer constraints on the bulk gas mass of each disk. In addition, further analysis of MWC 480 by \citet{Andrews_2024_mwc_mass} found kinematic mass of 0.13 $^{+0.04}_{-0.01}$ M$_{\odot}$, consistent to the value in this work. The consistency in our constraints on the bulk gas mass reinforce the estimations of the overall H$_{2}$ content assumed in our models. However, the kinematically derived gas masses do not provide direct constraints on the gas density profile interior to 50 au, so they alone cannot rule out an elevated inner disk gas column.

 Although constraining the gas surface density profile in the inner 50 au of these disks is challenging, observations of molecules other than CO can test whether the gas column as a whole rises steeply inside 50 au. In this scenario we would expect that many other molecular emission display strongly centrally peaked profiles similar to CO. \citet{Law_2021} presented high resolution line observations of HCN, DCN, HC$_{3}$N,CH$_{3}$N, C$_{2}$H, H$_{2}$CO, and c-CH$_{3}$N in HD 163926 and MWC 480. From Figures 22 and 23 of \citet{Law_2021}, it can be seen for HD 163296 only CO and its isotopologues show a centrally peaked emission profile. In contrast, HCN, DCN, C$_{2}$H, and c-CH$_{3}$N line emission all showed a central dip, while H$_{2}$CO and HC$_{3}$N had a central hole in their intensity profile. Similarly in MWC 480, C$_{2}$H also displayed a central hole feature, while HCN and HC$_{3}$N show a central peak in their emission.
 
 Given the diversity of radial profiles of molecular emission and the absence of central peaks in many other molecules, we argue that it is more likely that the steep increase in CO column density is due to an increase in CO abundance rather than an increase in the total gas content in the inner disk. Furthermore, \citet{Bosman_2021} compared the column densities of C$_{2}$H and C$_{18}$O for HD 163296 and MWC 480 derived by \citet{Zhang_2021}. In their analysis it can be seen that the column density of  C$^{18}$O increases inside 50 au while the column density of C$_{2}$H in the inner 50 au decreases towards the central star in both disks. Therefore it is less likely that the total gas column interior to 50 au has a steep increase, rather, CO is likely uniquely enhanced. In the future, measuring the line width due to pressure broadening can directly test whether the gas surface density increases steeply inside 50\,au \citep{Yoshida_2022}.  

In addition to comparing the kinematic masses and emission morphology of other molecules, we assess the gas disk mass of HD 163296 using HD (1-0) line flux upper limits determined by \citet{Kama_HD_2020}. They found an upper limit flux of 67 Jy km/s for the HD (1-0) line emission resulting from a non-detection of HD. Adopting methodology from \citet{Zhang_2020}, we produce two models of HD (1-0) line emission: one enhancing the HD content to the same degree as $^{13}$C$^{18}$O, representing an overall gas enhancement, while the other kept the HD abundance at the standard 2$\times$10$^{-5}$ ISM level. The results of each model is shown in Figure \ref{fig:HD_Abundance} in Appendix \ref{app:HD10}. We find the enhanced HD model had a total flux of 71 Jy km/s, while the unenhanced model had a total flux of 63 Jy km/s, within the upper limit. While the unenhanced model satisfies the observationally derived upper limit, the difference between the two models is not significant enough to rule out additional H$_{2}$ in the inner disk. Still, this analysis provides useful predictions to the previous upper limit presented by \citet{Kama_HD_2020}. Even the HD line flux does not provide strong constraints on the inner disk gas mass,  the kinematically derived masses and molecular emission profiles point to the enhancement of CO inside 50\,au for the HD 163296 disk. MWC 480 does not have measurements of HD line fluxes, and therefore we do not perform similar models.


\section{Discussion}{\label{sec:discussion}}

\subsection{Cumulative Pebble Drift}{\label{sec:Pebble_Flux}}

\subsubsection{Deriving Cumulative Pebble Mass}{\label{sec:deriving_cumulative_mass}}

The CO enhancement profiles shown in Figure \ref{fig:best_models} display centrally peaked enhancement of CO, suggesting significant pebble drift has occurred. We use these CO enhancement profiles to then estimate the total excess amount of CO gas interior to the snowline of each disk. We define excess CO gas as the leftover mass following the subtraction of CO gas that would be expected from purely ISM CO abundances. From the excess CO gas mass, we can then derive estimations for the total pebble flux required to supply the additional CO gas interior to the CO snowline. 

We begin this process by first calculating the gas surface density of both disks using equation \ref{selfsimilar} described in section \ref{gas_surface_density}
For our calculations, we again adopt the characteristic radius R$_{c}$, gas surface density exponent $\gamma$, and surface density at R$_c$, $\Sigma_c$, from Table 2 of \citet{Zhang_2021}. We integrate this equation throughout the disk and scale the CO abundance relative to an ISM ratio of CO/H$_{2}$ = 1.4 $\times$ 10$^{-4}$ by our derived CO enhancement factor for $^{13}$C$^{18}$O. We choose these profiles specifically as they probe deepest into the midplane, where CO would be actively sublimating off drifting pebbles. Additionally, we calculate the CO mass exclusively for radii interior to the CO snowline, as it then establishes a baseline for the excess CO gas resulting from pebble drift. We subtract the mass of CO gas resulting from the integration of equation \ref{selfsimilar}, but accounting for exclusively ISM carbon abundances to find just the excess CO gas mass interior to the CO snowline.  

Next, we assume that the total excess CO gas mass is due to the sublimation of CO ice from pebbles that were originally located outside the CO snowline and migrated inwards.
To estimate the total pebble masses that have drifted through the CO snowline, we assume that the pebbles will have compositions similar to comets. We adopt a refractory-to-ice mass ratio $3 \leq  \delta \leq 4.5$  and and all-ices-to-CO-ice mass ratio of 4 for our pebble composition \citep{Fulle_2019,2020A&A...636L...3F,Choukroun_2020}. This then gives us an estimate that CO ice will make up 4.5-6.25\% of the total pebble mass fraction. Using this mass fraction, we then provide an estimate of the total pebble mass within the inner disk. We note that the cometary abundances assumed are dependent on the formation environment of the comets, and pebble composition may vary across disks with differing CO ice abundances. In cases where the ice-to-refractory mass ratio decreases, that would lower our overall expected cumulative pebble mass estimate. In addition, our estimate serves as a lower limit on the pebble drift in each of these disks as we are not considering processes that result in CO destruction or processing. 

 \subsubsection{Cumulative Pebble Mass of HD 163296 and MWC 480} 
 
Understanding the cumulative pebble mass, being the total mass of pebbles that have drifted across the CO snowline, is crucial in assessing the pebble accretion models and the possible planet formation pathways of protoplanetary disks. Here, we use the methods described in section \ref{sec:deriving_cumulative_mass} to give an estimate for the total cumulative pebble flux in each of our disks. It is important to note that these calculations assume that the CO enhancement interior to the CO snowline is entirely due to volatile sublimation via pebble drift. 

We first look at the CO enhancement of the HD 163296 protoplanetary disk, which has CO abundances up to 15$\times$ ISM levels. Considering the entire CO gas content interior to the CO snowline, there is an enhancement of roughly 16 M$_{\oplus}$ of CO gas relative to what would be found with ISM abundances of CO. This requires a cumulative pebble mass of 250-350 M$_{\oplus}$ to have drifted interior to the CO snowline to produce such an enhancement. Our results remain consistent with previous literature that predicted a cumulative pebble mass of 150-600 M$_{\oplus}$ \citep{Zhang_2020}.

Turning to the MWC 480 protoplanetary disk, the CO enhancement profile displayed similar features to HD 163296, with centrally peaked CO enhancement of up to 10$\times$ ISM levels. We find that interior to the CO snowline, there is an excess of 30 M$_{\oplus}$ of CO gas. Again, adopting comet compositions for our pebbles, we find 480-660 M$_{\oplus}$ of pebbles are required to have drifted into the inner disk to produce the observed levels of CO enhancement.

\subsection{Cumulative Pebble Mass in Context of Planet Formation Models}{\label{sec:Planet_formation}}

A strong motivation for exploring pebble drift is to determine its influence on the planetary systems that may form from the influx of pebbles into the inner disk. In recent years, simulations have been performed with differing degrees of pebble flux to place thresholds on the cumulative pebble mass necessary to form different types of planets in the inner disk \citep[e.g.][]{Bitsch_2019,refId0}. Pebble accretion is believed to be most important in forming gas giants in the lifetime of the gas disk, with pebble drift playing a vital role in delivering solid material. \citet{Bitsch_2019,refId0} found that cumulative pebble masses of $\sim$200 M$_{\oplus}$  are sufficient to drive the growth of gas giants. \citet{Bitsch_2019} also found that if there is significant migration of planets within the disk, up to 350 M$_{\oplus}$ may be necessary to form gas giants. Using these mass thresholds, we can contextualize our calculated cumulative pebble masses for HD 163296 and MWC 480 in terms of the systems they may be able to form.

Looking first at the HD 163296 disk, we calculated a cumulative pebble mass ranging from 250-350 M$_{\oplus}$ throughout its lifetime ($\sim$5 Myr). In the upper limit of this regime, gas giants should be expected to form in all cases of planet migration. In the lower limit of our estimated mass flux, it is still possible to expect gas giants cores to form if planetary embryos do not experience significant rates of migration. This was detailed in \citet{Bitsch_2019} in the 200 M$_{\oplus}$ threshold, which is completely exceeded by our predicted mass range. Recent work looking into the structure of the HD 163296 disk would be consistent with the 200 M$_{\oplus}$ threshold as it is predicted that 2 sub-saturns have formed interior to the CO snowline \citep{Garrido-Deutelmoser_2023}. This also increases the confidence of our pebble mass estimations as they meet the necessary minimum masses to begin forming these types of planets.

We now move to our analysis of the MWC 480 disk where we found a cumulative pebble flux of 480-660 M$_{\oplus}$. Even in the lower limit of our cumulative pebble drift estimates, we predict that enough pebbles have drifted interior to the CO snowline to be capable of developing gas giants in the limits placed by both \citet{Bitsch_2019,refId0}. Our results are consistent with the current literature, as analysis of the dust structure of MWC 480 performed by \citet{Liu_2019} finds that a 2.3 Jupiter mass planet can be formed interior to 100 AU in MWC 480. This would be expected from our predicted pebble flux of 480-660 M$_{\oplus}$ compared to both gas giant formation thresholds of 200 M$_{\oplus}$ and 350 M$_{\oplus}$.

\subsection{
CO Enhancement Profiles and Pebble Drift}\label{sec:Pebble_Drift_Behaviour}

In our analysis, we also aim to constrain the behavior of pebble drift via our CO enhancement profiles. We look for distinct morphologies in the CO enhancement profile that were first described by \citet{Cuzzi&Zahnle_2004} and expanded simulations performed by \citet{Stammler_2017}, \citet{Krijt_2018}, and \citet{Krijt_2020}. These works showed that in the case of active pebble drift, a localized bump of CO enhancement would appear centered around the CO snowline as the icy mantles of the pebbles sublimated. Then it was proposed over longer timescales, in a steady state of pebble drift could result in a uniform enhancement of gas phase volatiles interior to the respective snowline \citep{Cuzzi&Zahnle_2004}. The third case of pebble drift outlines a scenario where the pebble flux is impeded due to the formation and build up of accreting objects in the path of the pebble flow. As a result, the volatile content is then depleted in the inner disk as the pebbles were accreted onto the objects.

\begin{figure*}
    \centering
   \includegraphics[width = 0.9\textwidth]{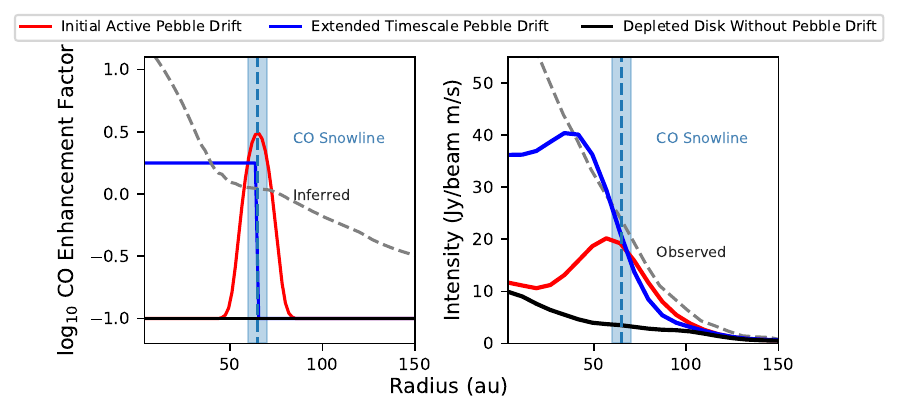}
    \caption{Three scenarios of pebble drift in the HD 163296 disk and model radial integrated profiles for the $^{13}$C$^{18}$O (2-1) line. CO enhancement factors are shown on a logarithmic scale (Left), representing no pebble drift, ongoing local pebble drift, and pebble drift occurring over long timescales and the resulting radial intensity profiles (right). In grey, we have also included the observed profiles for reference.}
    \label{fig:pebble_profiles}
\end{figure*}

To visualize the impacts of pebble drift, we implemented three toy models incorporating simple cases of pebble drift to generate CO enhancement and resulting radial intensity profiles for the $^{13}$C$^{18}$O (2-1) line using the HD 163296 disk as a representative disk. Each of these models were made using \texttt{RADMC3D}, and are shown in Figure \ref{fig:pebble_profiles}. We begin by displaying a disk with no CO enhancement by normalizing the enhancement profile to the CO abundances observed beyond the CO snowline \citep{Zhang_2021}. This scenario then serves as our baseline for a disk without any pebble drift. When producing model observations, we see a weak radial intensity profile that is slightly centrally-peaked due to the higher temperatures in the inner 50 au. To model active pebble drift, we inject a Gaussian centered around the CO snowline in the CO enhancement profile. This Gaussian was then reflected in the resulting radial intensity profile. Finally, to display a long history of pebble drift, we uniformly enhanced the disk with CO interior to the CO snowline. The corresponding radial intensity profile shown in blue is then much stronger interior to the CO snowline.

If we now look to Figure \ref{fig:best_models}, in order to reproduce the radial intensity profiles that were observed, we find that both disks need centrally peaked enhancement of CO. In addition to central enhancement, the regions closest to the CO snowline are depleted. We hypothesize that the centrally peaked CO enhancement is the result of significant pebble drift early in the lifetime of the disk. Due to the ages of these disks, we would expect pebble drift to have likely diminished in part due to less material and the presence of substructures in the disks, as simulations done by \citet{Krijt_2018} predict a majority of pebble drift occurs within the first 1 Myr of a disk's lifetime. In addition to their ages, both disks now have developed substructures which can also slow down the inward drift of pebbles \citep{Kalyaan_2023,Mah_2024,Krijt_Banzatti_2025}.
We do note however, that our derived CO enhancement profiles do not match the behaviors predicted in previous pebble drift models, as \citet{Cuzzi&Zahnle_2004} predicted extended pebble drift would produce a uniformly enhanced inner disk. In addition, we do not see local vapor-phase abundance enhancements just within the snowline, which has been predicted for active pebble drift \citep[e.g.][]{Booth_2017, Stammler_2017, Krijt_2018, Booth&Ilee_2019}. In contrast, the highest concentrations of CO are in the inner au. A contributing factor to this behavior and the significant mass of pebbles we have calculated to have drifted into each disks' snowline is the low turbulence observed in each disk. \citet{Flaherty_2020,Flaherty_2015} determined that the upper limits of the turbulence parameters $\alpha$ for the HD 163296 and MWC 480 disks were $\alpha < 9.6 \times 10^{-4}$ and $\alpha < 6 \times 10^{-3}$ respectively. In these regimes, it is expected that due to the weaker turbulence, factors such as gas advection would be weaker, while the inward drifting of pebbles would be enhanced \citep{Schneider&Bitsch_2021_chemcomp}.

It is also possible for the turbulence to be variable within in a disk rather than as a fixed $\alpha$ value, especially closer to the star. Due to this it is possible that gas advection could be heightened further from the star and begin to pile up within 50 au due to a steadily decreasing $\alpha$. In this case there would be an increased CO column density, along with increased hydrogen content in the inner disk. For the HD 163296 disk it was presented in sections \ref{Robustness} and Appendix \ref{app:HD10}
that there is a lack of an increased hydrogen density close into the star. While we could not do similar tests for MWC 480, we can look to the best-fit gas surface density to determine whether there is a significant decreasing alpha coefficient. Simulations have found that when the $\alpha$ parameter decreases inside 50 au, a sharp increase in the gas surface density can be seen. However, neither of our disks shows such a feature in our best-fit gas surface densities, rather there is a steady rise towards the center of the disk that would not cause a significant increase in CO column density as found in our models \citep{Delage_2023_Deadzone,Tong&Alexander_2025_Deadzone}. Thus we can be confident that the CO enhancement we have found is due to pebble drift, rather than turbulence driven effects.

In the following subsection, we implement more complex models to showcase the impacts that differing levels of turbulence have on the pebble flux and CO enhancement within a disk. In addition, we also consider additional factors that could drive the centrally peaked enhancement of CO, while being depleted around the CO snowline.

\subsection{Pebble Drift - Modeling Centrally Peaked CO}  
In this subsection, we explore two ideas on how to deliver CO-icy pebbles within the CO snowline to create a central peak in the context of the pebble drift model \texttt{chemcomp}\footnote{\texttt{https://github.com/AaronDavidSchneider/chemcomp}} \citep{Schneider&Bitsch_2021_chemcomp}. \texttt{chemcomp} is a 1D semi-analytical viscous disk model \citep{Lynden-Bell&Pringle_1974} containing several key mechanisms, such as dust growth, pebble drift, volatile condensation and sublimation, and gas advection and diffusion. It contains a basic chemical partitioning model, which traces the transport and distribution of several select volatiles, including CO and water. In this model, it does not track active chemistry, but rather considers the gas phase vs solid phase of the volatiles.


\subsubsection{Volatile Desorption Distances}\label{sec:Volatile_Desorption_Distances}

Small CO-ice-covered grains will sublimate almost instantaneously when they move to regions with $T>20~\rm{K}$. For larger-sized grains, however, the distance from the star at which these icy pebbles will sublimate may differ compared to smaller pebbles \citep{Piso_2015}. This desorption distance leads to pebbles penetrating further into the disk before fully desorbing and releasing CO. This is primarily due to larger pebbles' drift speed being higher than smaller pebbles, along with longer desorption timescales due to having less surface area per unit mass \citep{Piso_2015}. Consequently, this could explain how the CO vapor reaches so far into the disk without requiring fast gas advection. Here we mimic this behaviour in \texttt{chemcomp} by updating the desorption distance from $1\times10^{-3}~\rm{AU}$ (fiducial model) to 10 and 30 AU, based on Fig.~2 of \citet{Piso_2015} for $\sim$cm-sized pebbles releasing CO. The size of pebbles within the HD 163296 disk is not particularly well-constrained, but previous works \citep[e.g.][]{Sierra_2021,Guidi_2022,Williams&Krijt_2025} indicate that the pebbles may be close to $\sim$cm sizes, so it is plausible that pebbles may travel further into the disk this way.

\subsubsection{Volatile Trapping}\label{sec:Volatile_Trapping}
An alternative to varying desorption distances is trapping the CO in water ice, which has been a known phenomenon for some time \citep{Bar-nun_1985} but has been receiving increasing attention, and has recently been observed by JWST in the shape of ice band absorption profiles \citep{Bergner_2024}, which motivates this approach. This occurs when some CO ice is trapped within pores in water ice, and is released when water undergoes a transition from amorphous to crystalline form \citep{Collings_2003,Burke&Brown_2010}. Volatile trapping has been shown to significantly impact disk chemical properties, such as the C/O ratio \citep{Ligterink_2024}. We model this mechanism by redistributing the CO budget of the disk into two species: regular CO and trapped CO. For 0\% trapping, all of the CO budget goes into regular CO, and 100\% trapping locks the whole CO budget into water ice. Entrapment allows for the water ice pebbles to transport CO farther into the disk more easily compared to e.g. gas advection moving the sublimated vapor from the CO snowline. This can create a central peak while avoiding gas diffusion smearing out the abundance profile. Here, we vary the trapping fraction $\mathcal{T}$ as 0\%, 30\% and 60\%.

The majority of CO trapped in water ice is not released with water desorption, but instead when water changes from amorphous to crystalline structure \citep{Collings_2003, Burke&Brown_2010}. Water desorption occurs at a temperature of a few tens of Kelvin higher than crystallization, which corresponds to a difference of less than 1 AU in our models. Hence, for simplicity, we model the release of the trapped CO with water desorption.

Our model for the HD 163296 disk uses the same parameters given in Table 2 of \citet{Zhang_2021}, and we vary the turbulence parameter $\alpha$ \citep{Shakura&Sunyaev_1973} from $10^{-4}$ to $10^{-3}$ to explore the effect of a lower- and higher-turbulence disk. These ranges are also motivated by the alpha values presented by \citet{Flaherty_2015, Flaherty_2017} In order 
to compare against our observations, we calculate a CO enhancement factor as:

\begin{equation}\label{eqn:chemcomp_enhancement_factor}
    E_{\rm{CO}} = \left( \frac{\Sigma_{\rm{CO}}(r,t)}{\Sigma_{\rm{gas}}(r,t)} \right) \cdot \left( \frac{\Sigma_{\rm{CO}}(r,t=0)}{\Sigma_{\rm{gas}}(r,t=0)} \right)^{-1} 
\end{equation}

\noindent where $\Sigma_{\rm{CO}}(r,t)$ is the vapor surface density of CO at position $r$ and time $t$, and $\Sigma_{\rm{gas}}(r,t)$ is the total gas surface density, including CO. We fix the initial $\rm{C}/\rm{H}$ abundance to that given in Table 1 of \citet{Schneider&Bitsch_2021_chemcomp}, and 26\% of the carbon budget going into CO. The remaining carbon budget sees 10\% in CO$_2$, 45\% in CH$_4$, and 19\% in pure carbon grains. This is to match the simulation $\rm{CO}/{H_2}$ abundance to the ISM value of $1.4\times10^{-4}$.

\subsection{Pebble Drift - Model Results Consistent with Volatile Trapping}

\begin{figure*}
    \centering
   \includegraphics[width=\textwidth]{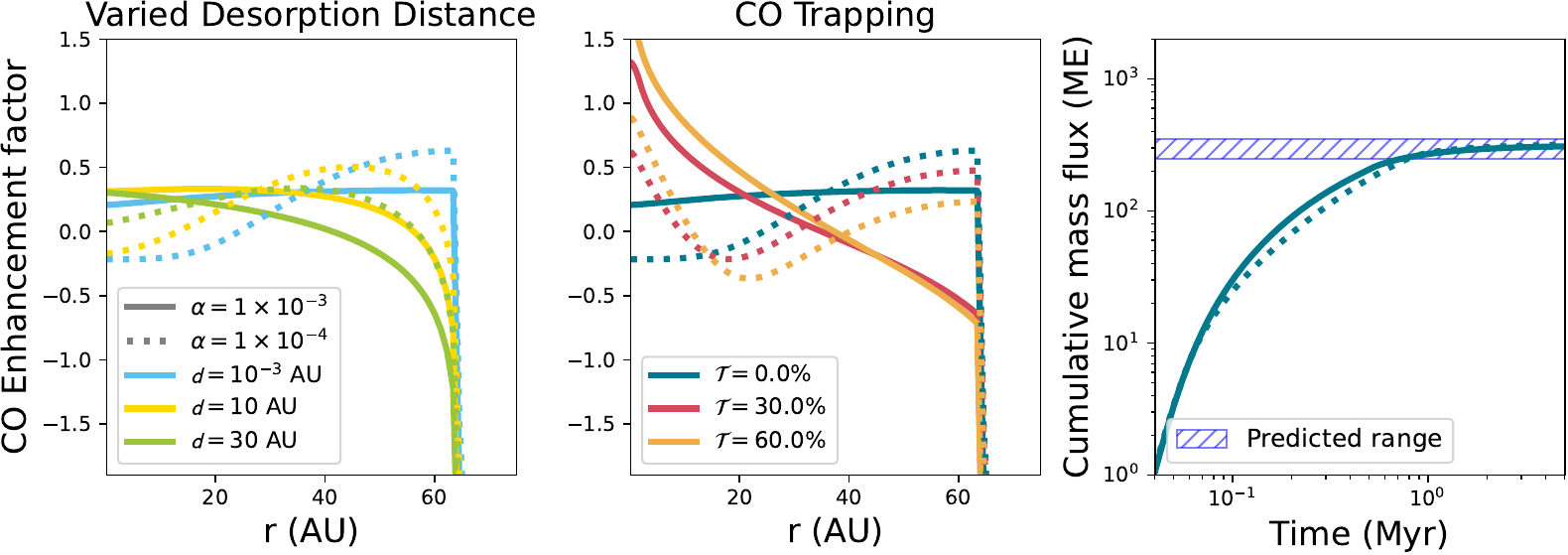}
    \caption{Radial profile of the logarithm of the CO enhancement factor (Eq.~\ref{eqn:chemcomp_enhancement_factor}) as calculated by the code \texttt{chemcomp} \citep{Schneider&Bitsch_2021_chemcomp} for a model of HD 163296, 5 Myr into the simulation. \textit{Left:} variation of the CO enhancement profile with desorption distance $d$, with longer distances permitting icy pebbles to reach further into the disc (see text for details). \textit{Middle:} variation of the enhancement profile with different trapping fractions of CO in water ice (0, 30, and 60\%). Each panel shows models of different turbulence values $\alpha$ in dashed and dotted lines. \textit{Right:} Cumulative mass flux of pebbles through the CO snowline for the two $\alpha$ models with no trapping and $d=1\times10^{-3}~\rm{AU}$, also featuring no CO and N$_2$ sublimation (preventing material recycling; see text). The models shown here fit comfortably within the mass delivery constraints from the observational data.}
    \label{fig:chemcomp_models}
\end{figure*}

Figure~\ref{fig:chemcomp_models} shows the logarithm of these CO enhancement profiles at $t=5~\rm{Myr}$ as calculated from the simulation outputs. Since the enhancement profile is defined as the abundance of CO in the gas phase with respect to the initial CO abundance at $t=0$, the enhancement factor at $t=0$ is a flat line at $y=0$. The left panel of 
Figure~\ref{fig:chemcomp_models} shows that increasing desorption distances to larger values (10 to 30 AU) provides an effective means of transporting CO to the inner disc; this is especially evident for the lower turbulence $\alpha=10^{-4}$ model. Higher turbulences (and so faster gas advection) deliver the CO to the inner disc more quickly, creating a depletion near the CO snowline and a comparative enhancement in the inner disk, as seen in the observational data. This result suggests that larger desorption distances (10 to 30 AU) require higher turbulences ($10^{-3}$) to explain a CO abundance higher than that at the CO snowline, as seen in the observations in Figure~\ref{fig:best_models}.

Volatile trapping, however, provides a better match to the observations for lower turbulence models ($10^{-4}$), and is best-fitting models at higher turbulences ($10^{-3}$), where the enhancement profile is similar in shape to the observations. Moderate-to-high trapping fractions (30 to 60\%) create a central peak in the inner disc, indicating that volatile trapping may provide a more accurate explanation for the centrally-peaked CO profile compared to traditional volatile transport models.

Comparing the net mass flux of pebbles against the expected value from observations is complicated due to the diffusion and re-condensation of vapor after sublimation around a snowline, causing pebbles to grow in a recycling fashion \citep{Ros&Johansen_2013}. This effect boosts the cumulative mass flux through the snowline by a factor of a few, making the simulation result initially incomparable. However, we can mitigate the enhancement of mass flux arising from re-condensation by restricting the sublimation of CO and N$_2$ ice, thereby preventing the recycling of vaporized volatiles.

Using this approach with the fiducial model ($d=10^{-3}~\rm{AU}$, $\mathcal{T}=0$\%, $\alpha=10^{-3}$), we find that $\sim$317\,$M_{\oplus}$ of pebbles is delivered to the CO snowline, which is consistent within the expected range of 250-350\,$M_{\oplus}$. We extend this to the water snowline by then repermitting CO and N$_2$ sublimation, but disallowing water sublimation. This model then leads to a mass flux of $\sim 308 M_{\oplus}$ through the water snowline, which is consistent with previous estimates for HD 163296 \citep{Williams&Krijt_2025}. 

Our results suggest volatile trapping is a plausible explanation for the observed CO enhancement, however, we note that there are a few caveats in our procedure. Our predicted mass flux is contingent on the chemical partitioning model used: we locked some portion of the $\rm{C}/\rm{H}$ budget into pure carbon grains instead of CO in our no-CO model, but it may instead go into e.g. $\rm{CH}_4$, which would sublimate before the water snowline. This would reduce the solid mass flux to the very inner disc, impacting our predictions.


This work is limited by the fact that HD 163296 is a substructured disk, whilst we have modeled it as a smooth disk. In the absence of gaps, the flow of pebbles remains unrestricted throughout the lifetime of the disk as gaps present pressure traps that can impede pebble drift \citep{Kalyaan_2021}. In addition, the location and time of formation of gaps within a disk can greatly affect the enrichment of vapor-phase volatiles within the inner regions of a disk, especially over long timescales as was shown by \citep{Kalyaan_2023,Krijt_Banzatti_2025}. Thus, in our models the CO enhancement could be decreased by the gap structure of HD 163296, however, the timescale of gap formation must also be considered. In Figure \ref{fig:chemcomp_models}, we see that a majority of the pebble flux occurs within the first Myr, and as HD 163296 has an estimated age of 4 Myr. If the gaps formed after the first Myr then it is possible that the CO enhancement profile would remain relatively unchanged. In the case that the gaps had formed earlier, it is possible that they would inhibit some pebble drift, or possibly prolong it. Simulation work by \citet{Easterwood_2024} found in disks containing two or three gaps, vapor enrichment at the water snowline can peak at 2 or 3 Myr due to delayed pebble drift rather than at 1 Myr. This could aid in creating the centrally peaked CO enhancement when considering a delayed release of CO caused by volatile trapping. However, due to the CO snowline being located at 60 au, the impact of gaps would be lessened for a majority of the CO that would be released. Therefore we believe that the enrichment seen is consistent more with early large scale pebble drift within 1 Myr, than late large scale drift. Future work will be necessary to determine the impact that gap structure has on the CO enhancement profiles of more extended disks. 

Another caveat to these results is that our assumed chemical partitioning model may, in reality, be different from what we have assumed here. Therefore more detailed future modeling is required to fully assess the impacts of volatile trapping. This idea has been explored in a more general disk scenario by \citep{Williams_2025_volatile_trapping}.
 
In summary, our models show that large pebbles are an effective means of explaining the centrally-peaked CO enhancement profile. This is because the pebbles can drift a considerable distance before sublimating CO ice, although trapping of CO inside water ice is more effective in reproducing the centrally-peaked CO enhancement profiles. For the cases considered here, higher $\alpha$ is preferable as $\alpha=10^{-4}$ causes a secondary peak at the CO snowline, which is absent in the observations (Fig.~\ref{fig:best_models}).

\subsection{Pebble Drift -  Persistent CO enhancement}

In the previous section we detail that the majority of the pebbles that entered the CO snowline at 70 au, drifted in within the first 1 Myr, with about 90\% of these pebbles reaching the water snowline at about 5 au over the 5 Myr. As 30\% of the CO is trapped in water ice, this pebble delivery was able to reproduce the CO enhancement profile we would expect at about 4-5 Myr. However, an important point to address in theis pebble delivery is the viscous timescale of this disk, as the gas will not remain static in the disk for the remaining 4 Myr after it had been delivered. We calculate the viscous timescale with the following equation:

\begin{equation}
    t_{vs} = \frac{R^2}{\nu} = \frac{R^2}{\alpha *c_s * H}
\end{equation}

To estimate the time the gas would persist assuming only inward advection at each snowline, we assume T = 20 K at 70 au and T = 150 K at 5 au, with H = 0.1 R, we obtain t$_{vs}$ $\approx$ 12.7 Myr at 70 au and $\approx$ 0.9 Myr at 5 au. In the case for the viscous timescale at 70 au, being much longer than the age of the disk this would allow for a rapid delivery of pebbles through the CO snowline, and a slower advection of gas consistent with our results. However at first glance, the 0.9 Myr viscous-timescale at 5 au would suggest that CO released there should be quickly advected inward and lost within a few Myr. However, this interpretation neglects outward diffusion: the CO released at 5 au both follows the inward viscous flow and diffuses outward, prolonging the persistence of the central CO enhancement. This affect is visible in the difference between the $\alpha$ = 10$^{-4}$ and $\alpha$ = 10$^{-3}$ cases shown in the central panel of Figure \ref{fig:chemcomp_models}.

Our 1D \texttt{chemcomp} models includes the competing processes of inward advection, outward diffusion, and viscous evolution ensuring their affects are fully captured. Taken together, they indicate that our pebble mass estimate and the resulting CO distribution are consistent with most pebbles crossing the CO snowline at about 70 au within the first Myr. In reality, both the HD 163296 and MWC 480 disks developed substructures within the first ~5 Myr. These rings and gaps can act as partial barriers to radial transport, slowing the inward flow of dust and gas. As a result, the substructures may further prolong the signature of enhanced CO abundances produced by pebble drift, allowing the signal to persist well beyond the simple viscous timescale.

\subsection{Discussion on CO Chemical Conversion}

The models used in this study hold that CO is the primary carrier of carbon compared to other species such as CO$_{2}$., which is assumed due to both MWC 480 and HD 163296 being Herbigs. In the case of disks around T-Tauri  there can be a significant depletion of CO gas due to chemical conversion into molecules such as CO$_{2}$ and CH$_{3}$OH \citep{Kama_2016_Volatile,Schwarz_2018_CO,Bosman_2018b_CO_Destruction}. However for Herbigs CO depletion is mitigated due to less chemical conversion and CO freeze-out caused by the hotter environment of Herbig disks \citep{Bosman_2018b_CO_Destruction,Oberg_2023_CO}. Additionally this effect was observed by \citet{Trapman_dynamic_disk_mass_2025}, who saw that among their sample, the Herbig disks had CO abundances very close to the ISM, whereas the T Tauri disks were much more depleted. This supports that there are different chemical pathways for CO processing among the T Tauri and Herbig disks, where around Herbigs a large fraction of the CO remains in the gas phase and show  little processing. Therefore our assumptions of locking 25\% of carbon in CO is reasonable due to less CO processing in Herbig disks.

\subsection{Dependence on Disk Temperature Structure}
In our analysis the CO abundances that we calculate are dependent on the gas temperature structure that we adopt for our modeling, and thus our pebble mass estimates are also dependent on this temperature structures. In section \ref{sec:methods}, we state that we adopt gas temperature structures developed by \citet{Zhang_2021}. These temperature structures have been shown to be consistent with empirical temperature structures derived by \citet{Law_2021_temperature}, using the emitting surface layers of CO and $^{13}$CO (2-1). From this analysis it was shown that temperature structures developed using thermochemical models are reliable within regions with temperatures of 15K-40K \citep{Zhang_2021}. 

However, recent work by \citet{Qi&Wilner_2024} has found another possible temperature structure that could explain the significant increase in CO column density interior to the CO snowline without requiring pebble drift and delayed desorption of CO ice. \citet{Qi&Wilner_2024} shows that a thick, vertically isothermal region around the midplane (VIRaM) can produce a sharp transition in CO column density interior to the snowline. An indicator of the VIRaM is a local minimum present in the derivative of the CO radial intensity profile. In Appendix \ref{app:Derivative} we present Figures \ref{fig:HD_derivative} and \ref{fig:MWC_derivative} which show the derivatives of the radial intensity profiles presented in Figure \ref{fig:best_models}. In Figure \ref{fig:HD_derivative} we see that a local minimum is present in the more optically thick lines C$^{17}$O and C$^{18}$O observed in the HD 163296  disk, consistent with \citet{Qi&Wilner_2024}. However, in the derivative of the $^{13}$C$^{18}$O J= 2-1 radial intensity profile, we see a potential local minimum, however with our uncertainties it is less robust than in the more optically thick lines. In addition, we did not find strong evidence for a local minimum in the derivatives in any of the CO lines observed in the MWC 480 disk, consistent with \citet{Qi&Wilner_2024}.  

The absence of strong detections of local minima around the CO snowline in the derivatives of the radial intensity profiles of each CO line suggests that a thick VIRaM layer may not be present in the temperature structure of the MWC 480 disk. In the case of HD 163296, a local minimum was present in the optically thick CO lines, but could not be verified within the $^{13}$C$^{18}$O observations with the current resolution, warranting future investigation at higher resolutions to determine the presence of a thick VIRaM layer. Due to this, we believe that the temperature structures derived from thermochemical models still accurately represent the region interior to the CO snowlines of each disk without including a thick VIRaM layer. Without a VIRaM layer, the resulting increase in CO column density would require significant CO enhancement resulting from pebble drift early on within the lifetimes of each disk.

\section{Conclusion}{\label{sec:conclusion}}
In this work, we present new ALMA observations of spatially resolved 0.3'' and 0.4'' resolution $^{13}$C$^{18}$O line emission observations of the HD 163296 and MWC 480 protoplanetary disks. Using radiative transfer and thermo-chemical models of these two disks, we provide the first spatially resolved CO enhancement profiles inside the CO snowlines of the two disks. Each disk showed significant centrally peaked  CO enhancement in the inner disk, exceeding 10$\times$ ISM abundance levels in the inner 20-30 AU region. Utilizing these results, we then estimate the pebble drift history for each disk. This analysis is summarized as follows:
\begin{enumerate}

    \item We estimate the respective CO enhancement factors for each disk where they showed centrally peaked CO enhancement while being depleted at the CO snowline. 

    \item The CO enhancement profiles provide an estimate of the excess CO gas interior to the CO snowline for both disks. The HD 163296 disk has an excess $\sim$ 16 M$_{\oplus}$ of CO gas, while the MWC 480 disk has $\sim$ 30 M$_{\oplus}$ of excess CO gas.

    \item We predict the total cumulative pebble masses required to reproduce the CO enhancement for each disk assuming the pebbles had comet-like composition. 250-350 M$_{\oplus}$ drifted within the CO snowline of HD 163296, while 480-660 M$_{\oplus}$ had drifted interior to the snowline of MWC 480.

    \item To explore how centrally CO enhancement can persist for millions of years as the result of an early history of pebble drift we present a model of the HD 163296 disk to serve as a general case. We used the 1D viscous disk pebble drift model \texttt{chemcomp} to track the pebble drift and CO enhancement while varying conditions such as desorption distance and volatile trapping of CO ice. From these models, we found that in turbulent disks, volatile trapping with efficiencies as low as 30\% mimicked the centrally peaked CO enhancements after 5 million years.

    \item Pebble drift is critical in the formation of giant planets in the lifetimes of the protoplanetary disks. Previous analysis of the MWC 480 and HD 163296 disks find that gas giants are expected to have formed interior to their respective CO snowlines, which would require a significant degree of pebbles to drive this formation. Our results remain consistent with these findings as we estimate sufficient cumulative pebble masses in both disks to form giant planets interior to the CO snowlines.

\end{enumerate}

In summary, our analysis of the radially dependent CO enhancement in HD 163296 and MWC 480 found that each disk displays a centrally peaked CO gas distribution much greater than the ISM abundances. Utilizing the CO gas distribution, we find that each disk requires cumulative pebble fluxes of hundreds of earth masses in order to reproduce the observed levels of CO. From our modeling of the HD 163296 disk, we find that volatile trapping may serve as a significant mechanism for producing a centrally peaked CO enhancement profile, with depletion starting towards the CO snowline. This builds upon the growing notion that volatile trapping could be critical for influencing the carbon abundances in the inner disk, along with the final C/O ratios observed in exoplanet atmospheres \citep{Ligterink_2024}. In the future, it will be essential to develop more complex 2-dimensional models of pebble drift and volatile trapping to better characterize the dynamics and long-term effects of pebble drift on the chemical contents within a protoplanetary disk.

\begin{acknowledgments}
This paper makes use of the following ALMA data: ADS/JAO.ALMA\#2021.1.00899.S. and ADS/JAO.ALMA\#2018.1.01055.L. ALMA is a partnership of ESO (representing its member states), NSF (USA) and NINS (Japan), together with NRC (Canada), MOST and ASIAA (Taiwan), and KASI (Republic of Korea), in cooperation with the Republic of Chile. The Joint ALMA Observatory is operated by ESO, AUI/NRAO and NAOJ. In addition, publications from NA authors must include the standard NRAO acknowledgement: The National Radio Astronomy Observatory is a facility of the National Science Foundation operated under cooperative agreement by Associated Universities, Inc. We would also like to thank our anonymous referee for their helpful feedback on improving the analysis in this work.

T.A., K.Z., and L.T. acknowledge the support of the NSF AAG grant \#2205617. J.W. is funded by the UK Science and Technology Facilities Council (STFC), grant code ST/Y509383/1. Support for C.J.L. was provided by NASA through the NASA Hubble Fellowship grant No. HST-HF2-51535.001-A awarded by the Space Telescope Science Institute, which is operated by the Association of Universities for Research in Astronomy, Inc., for NASA, under contract NAS5-26555. F.A. is funded by the European Union (ERC, UNVEIL, 101076613). Views and opinions expressed are however those of the author(s) only and do not necessarily reflect those of the European Union or the European Research Council. Neither the European Union nor the granting authority can be held responsible for them.

\end{acknowledgments}

\bibliography{sample631}{}
\bibliographystyle{aasjournal}

\appendix
\section{Additional Modeling Plots}\label{appendix:A}

\begin{figure*}
    \centering
    \includegraphics[width=\textwidth]{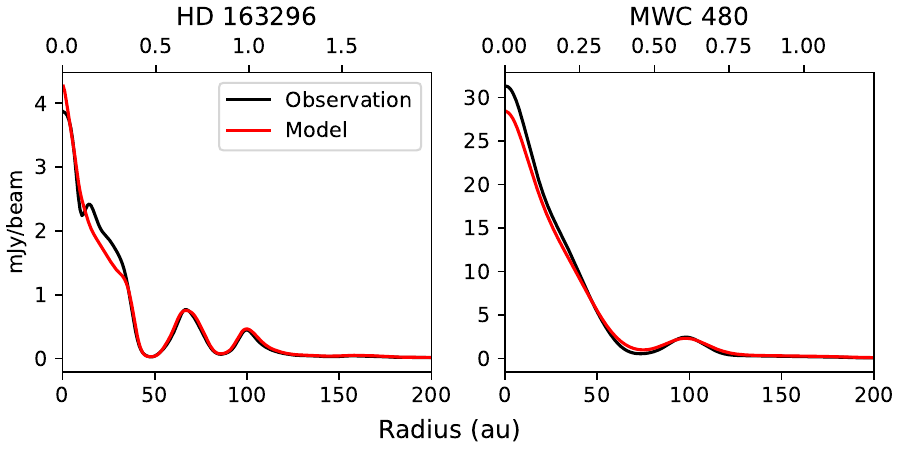}
    \caption{Show above is the dust continuum radial profiles for the HD 163296 and MWC 480 disks. Shown in red is our best-fit model generated while considering anisotropic photon scattering, compared to the observations shown in black. We note that there is some under-prediction in our models due to the optical depth of the dust. However, this is not significant enough to affect our results. }
    \label{fig:dust_radial_profile}
\end{figure*}

\begin{figure*}
    \centering
    \includegraphics[width=\textwidth]{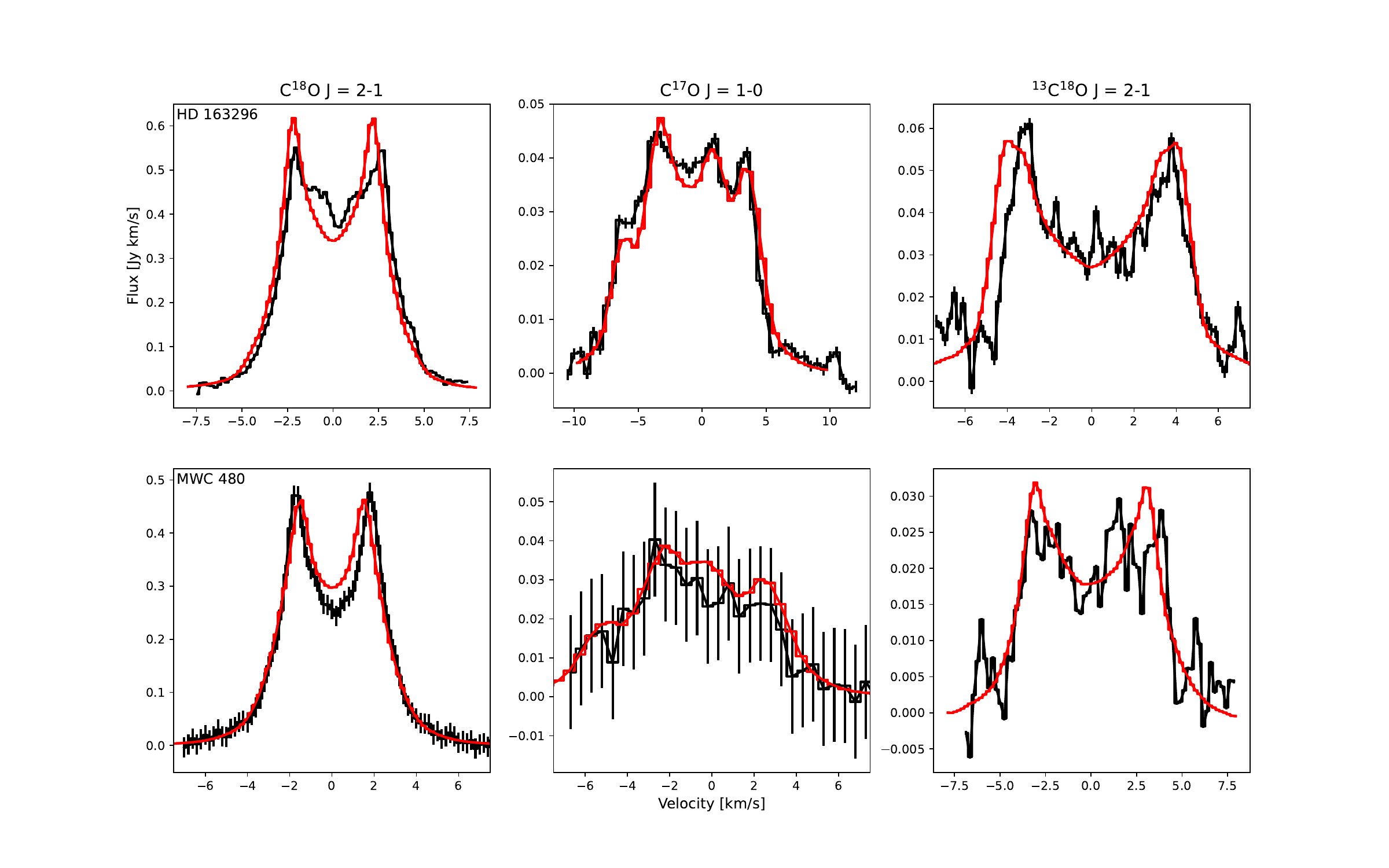}
    \caption{Top: Observed integrated spectra for HD 163296 along with the best-fit model over plotted in red for he t$^{13}$C$^{18}$O J =2-1, C$^{17}$O J = 1-0, and C$^{18}$O J = 2-1 emission. Bottom: Displayed is the observed spectrum annd best-fit models for MWC 480.}
    \label{fig:Non-projected_spectrum}
\end{figure*}

\begin{figure*}
    \centering
    \includegraphics[width=\textwidth]{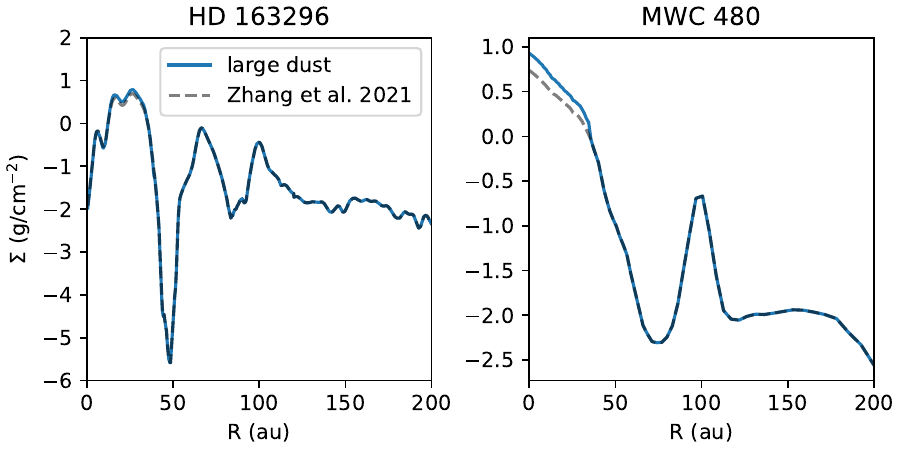}
    \caption{Shown above is the best-fit surface density profile for the large dust grain populations used in our models. Left: The large grain surface density profile for the HD 163296 disk. Right: The large grain surface density profile for the  MWC 480 disk. In blue is our best-fit surface density, while in gray is the best-fit surface density presented in \citet{Zhang_2021}}
    \label{fig:surface_density}
\end{figure*}

\begin{figure*}
    \centering
    \includegraphics[width=\textwidth]{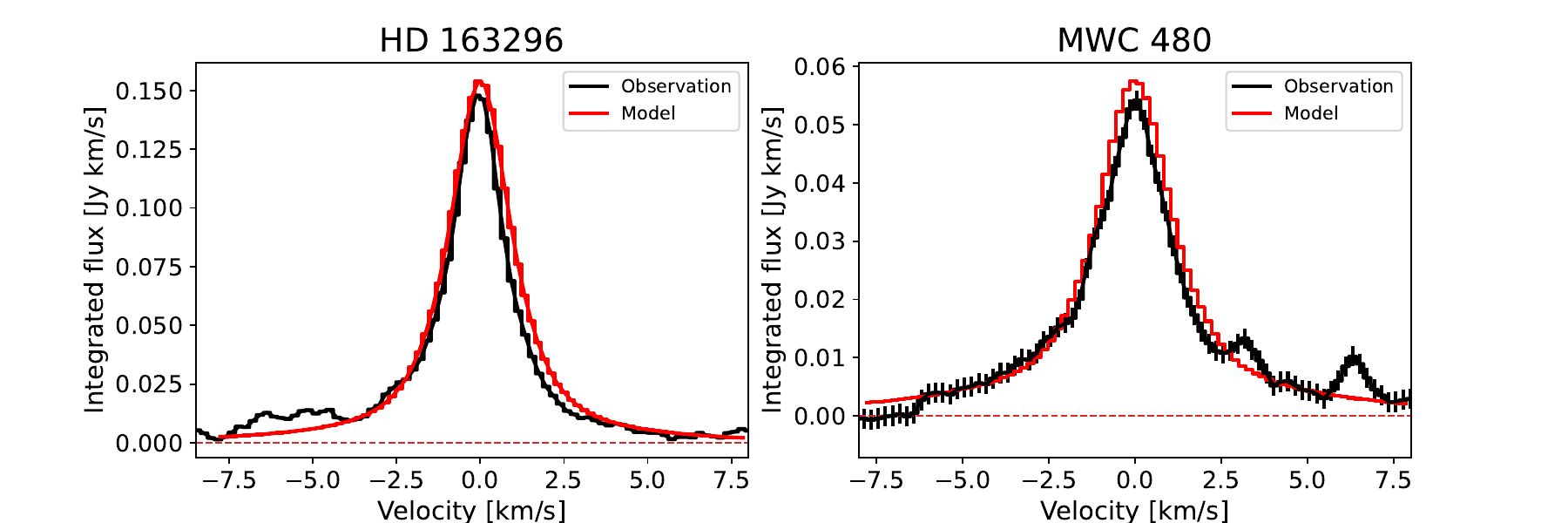}
    \caption{Displayed is the observed projected integrated spectrum (black) and our best-fit models (red) for the $^{13}$C$^{18}$O J = 2-1 emission of HD 163296 (Right) and MWC 480 (Left). We note that for MWC 480 disk our model overestimates the projected spectrum in the region of $\pm$ 1 km/s. This is likely due to a slight overestimation in the outer disk region, where the noise level and the signal strength become comparable. }
    \label{fig:spectra}
\end{figure*}

In this section, we include additional plots mentioned in Section \ref{sec:methods} of our best fit models. Figure \ref{fig:dust_radial_profile} shows our best-fit dust continuum radial profile compared to that of the 1.3mm dust continuum observations. Then, Figure \ref{fig:surface_density} shows our updated best-fit surface density profile of the large dust grain population used in our models. Figure \ref{fig:spectra} then shows the observed projected integrated spectrum and our best-fit models for both HD 163296 and MWC 480 $^{13}$C$^{18}$O observations. In addition to the projected spectra we have also included non-projected spectra along with the best fit model for each isopotologue. As can be seen in Figure \ref{fig:Non-projected_spectrum}, in each case there is wide broadening of the emission, indicating that a strong portion is coming from the enhanced inner disk region. 

\section{Comparison of Scattering and No Scattering Models}

\begin{figure*}
    \centering
    \includegraphics[width=\textwidth]{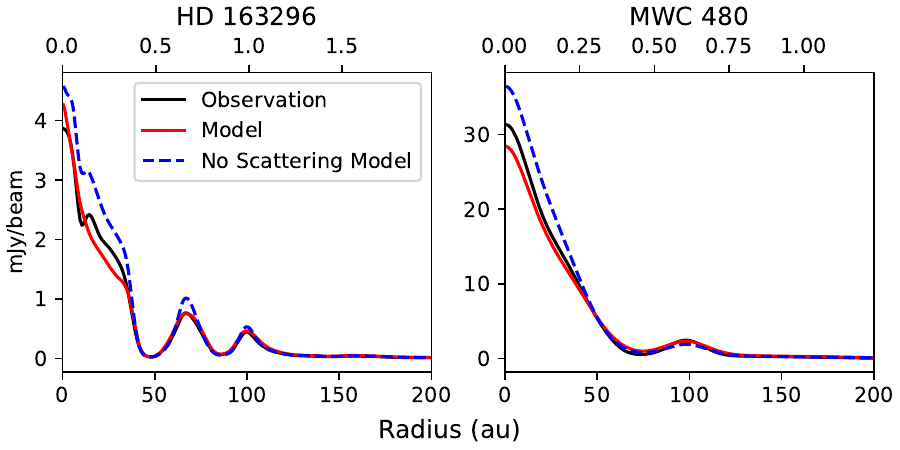}
    \caption{Shown above is a comparison of radial profiles of the dust continuum for HD 163296 (Left) and MWC 480 (Right), generated with and without anisotropic scattering. In blue, the model is without anisotropic scattering, while in red, the model is with scattering. In black is the radial profile directly from the dust continuum observations.}
    \label{fig:scattering_comparison}
\end{figure*}

In this section, we show a comparison of our dust continuum modeling when incorporating anisotropic scattering and ignoring scattering. This is best displayed in Figure \ref{fig:scattering_comparison}, where we can see that for the innermost regions of both disks, the models without scattering over-predict the observed emission. This is directly in line with what we would expect from \citet{Zhu_2019}, where the dust is expected to become optically thick, as our models with scattering accurately account for the increased impacts from the dust. The no-scattering model also displays a gap that is not present when scattering is incorporated. This is due to the additional Monte Carlo noise when accounting for the scattering effects from the optically thick dust in this region. 

\section{Derivatives of the Radial Integrated Intensity Profile}\label{app:Derivative}

In this section, we provide two figures showing the derivative of the observed radial intensity profiles for the MWC 480 and HD 163296 disks. In each figure we see that in the optically thin line there is not a strong local minimum  near the CO snowline. This indicates that there is likely not a vertically isothermal region around the midplane layer in the temperature structure of these disks.

\begin{figure*}
    \centering
   \includegraphics[width=\textwidth]{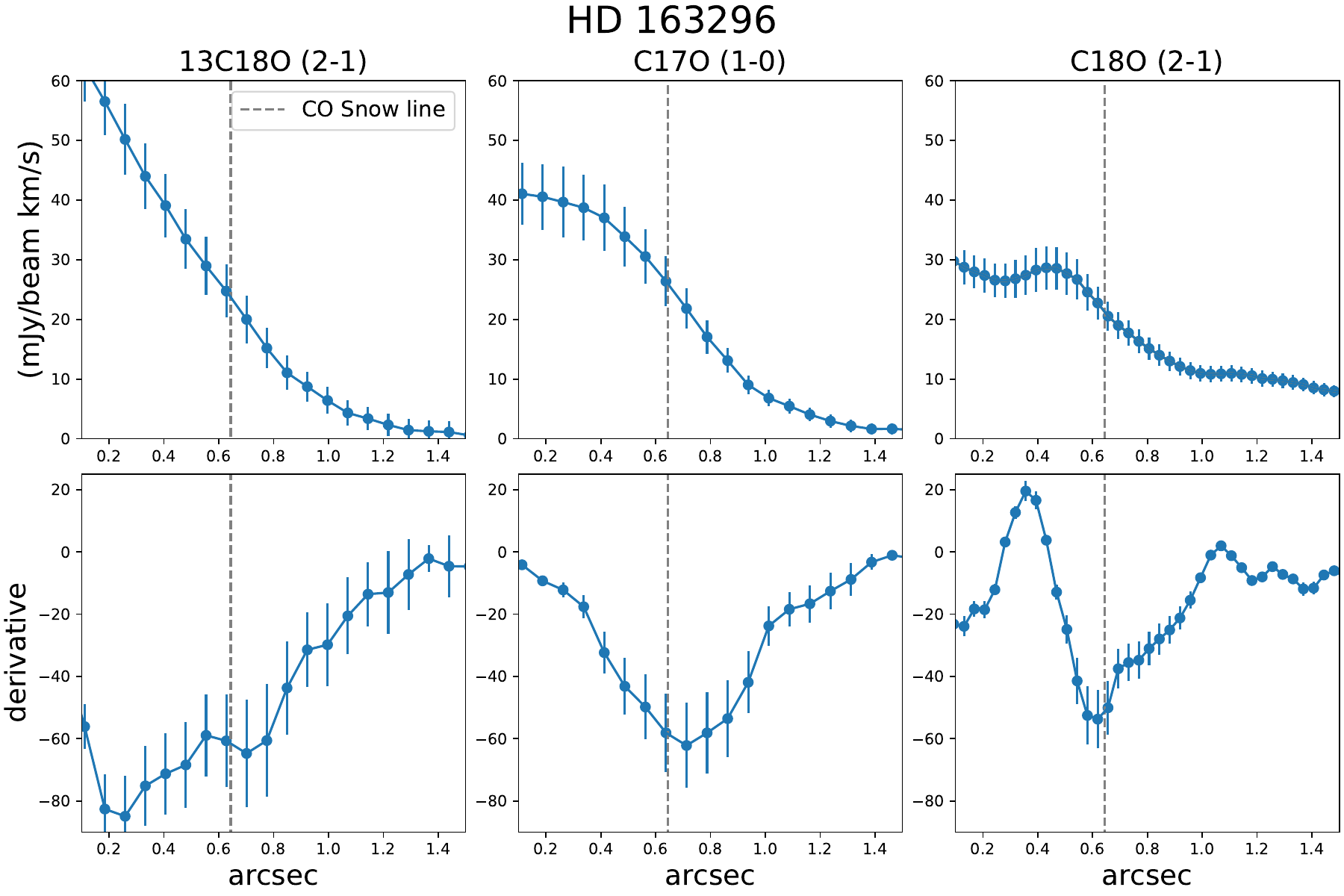}
    \caption{Top: The observed radial integrated intensity profiles for the CO isotopologues analyzed in the HD 163296 disk, also shown in Figure \ref{fig:best_models}. Bottom: Plotted are the derivatives of each of the respective radial intensity profiles, along with the CO snowline location at 65 AU as used in this study.}
    \label{fig:HD_derivative}
\end{figure*}

\begin{figure*}
    \centering
   \includegraphics[width=\textwidth]{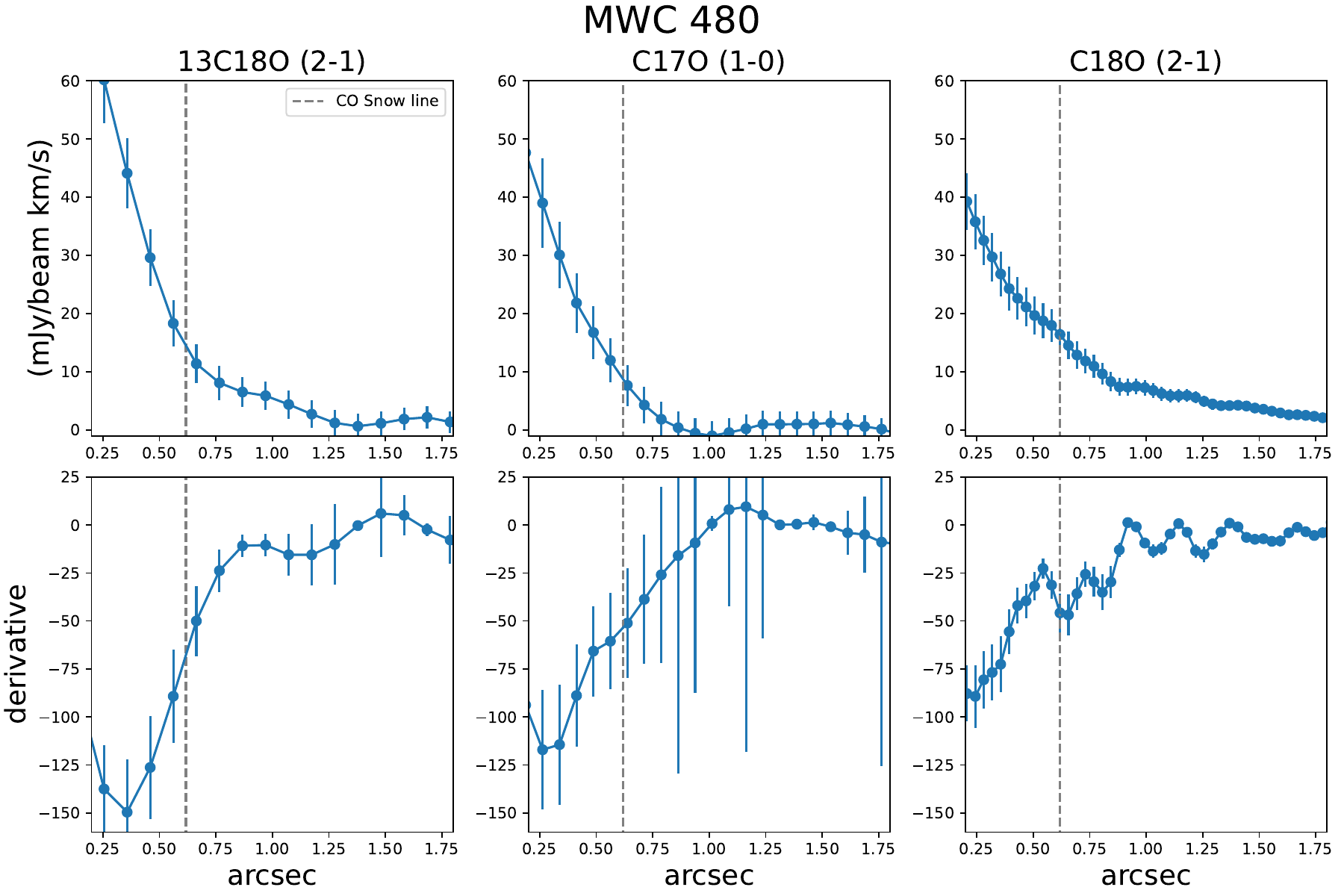}
    \caption{Top: The observed radial integrated intensity profiles for the CO isotopologues analyzed in the MWC 480 disk, also shown in Figure \ref{fig:best_models}. Bottom: Plotted are the derivatives of each of the respective radial intensity profiles, along with the CO snowline location at 100 AU as used in this study.}
    \label{fig:MWC_derivative}
\end{figure*}

\section{HD (1-0) Line Emission Modeling}\label{app:HD10}

\begin{figure*}
    \centering
   \includegraphics[width=\textwidth]{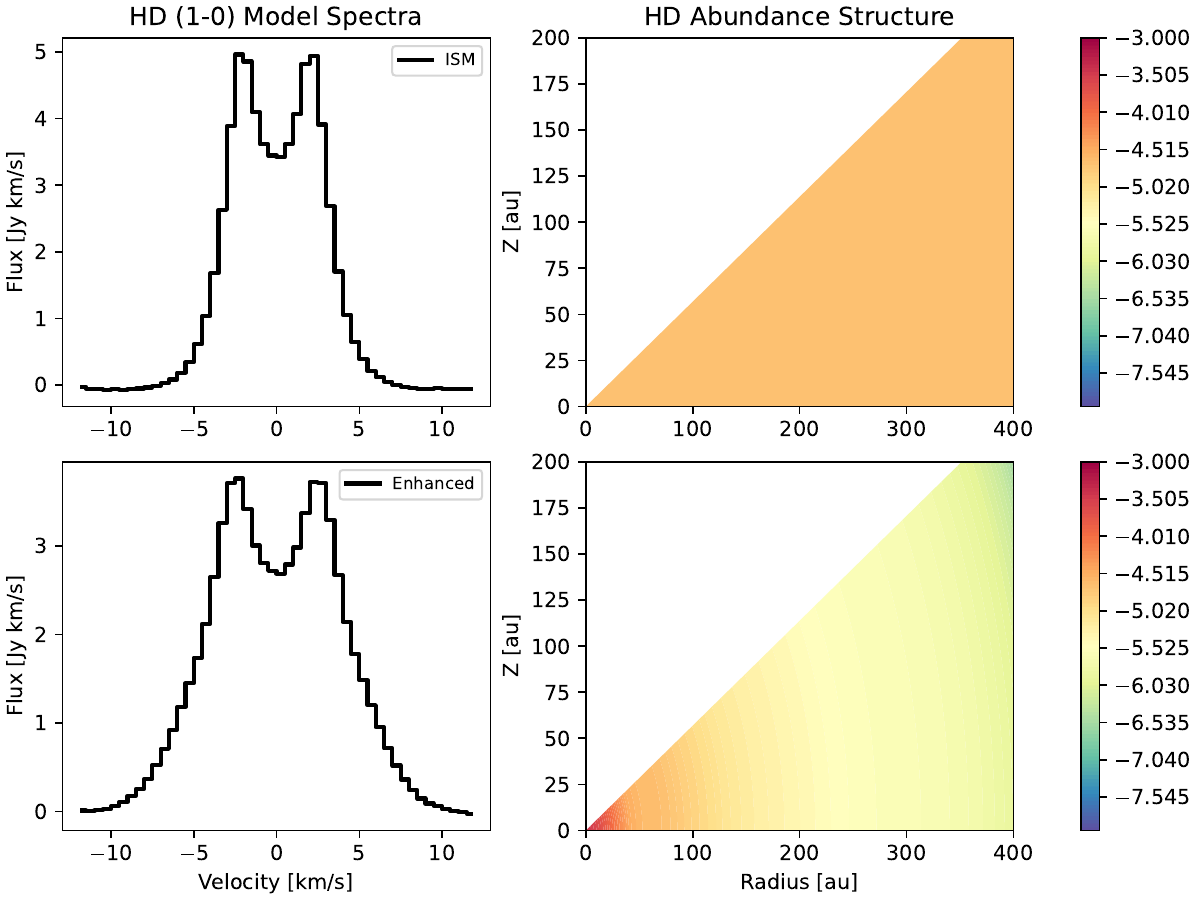}\caption{Top: These figures display a scenario where the HD content in our models has not been enhanced, and instead is set to typical ISM abundances of HD/H$_{2}$ = 2$\times$10$^{-5}$. The left hand displays the resulting model spectra, while the right shows the 2-D abundance structure of the HD  gas content. Bottom: These plots display the spectra and 2-D abundance structure of the HD content in our HD 163296 model after the same enhancement factor of $^{13}$C$^{18}$O has been applied. }
    \label{fig:HD_Abundance}
\end{figure*}

Here we show the results of modeling the HD (1-0) line emission in HD 163296 in Figure \ref{fig:HD_Abundance}. The methodology used to reproduce these line images is the same as described for the CO modeling in section \ref{Abundance Modeling}. The initial abundance was developed by multiplying the H$_{2}$ content in the disk by a factor of 2$\times$10$^{-5}$ to reflect the ISM ratio of HD/H$_{2}$. We then summed the spectral emission of each model and compared them to upper limits on the HD (1-0) emission found by \citet{Kama_HD_2020}. When summed the unenhanced model reaches a total flux of 63 Jy km/s, within the upper limit of 67 Jy km/s. 

\section{Effect of Beam Dilution on Models}

\begin{figure*}
    \centering
   \includegraphics[width=\textwidth]{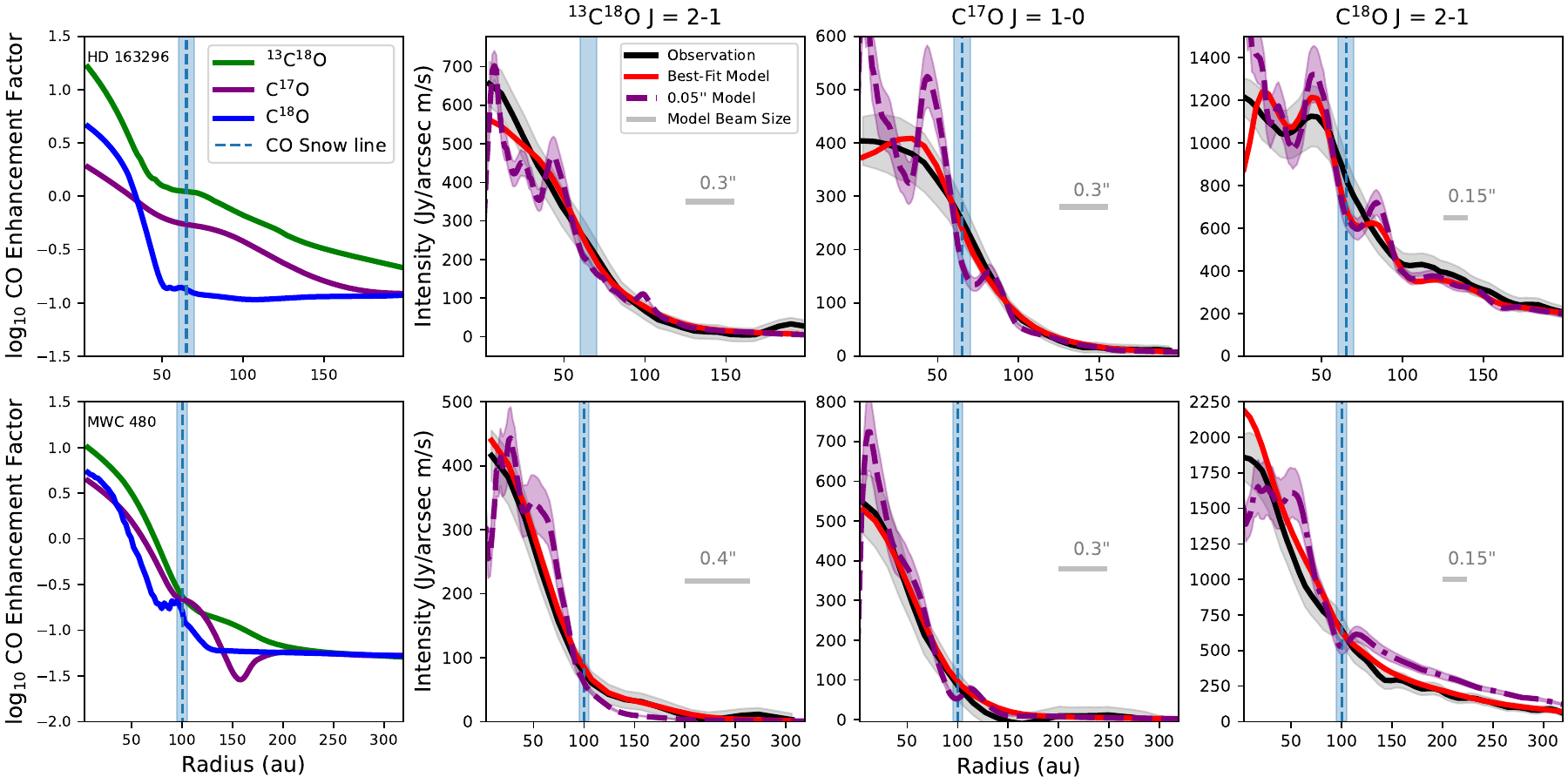}\caption{This figure displays the same information as Figure \ref{fig:best_models}, with the addition of model radial intensity profiles convolved to a 0.05'' beam. In addition the intensities have been converted to Jy/arcsec rather than Jy/beam.}
    \label{fig:High_Resolution_Models}
\end{figure*}

As part of our modeling procedure, when generating line image cubes we convolve the model with the same beam size used in observations. Due to this it is possible that some local maxima in the CO gas emission may be smoothed out due to the resolution. To demonstrate the effect of the beam we have generated additional higher resolution models for each isotopologue displayed in Figure \ref{fig:best_models}. These models were convolved to 0.05'' resolution and are shown in Figure \ref{fig:High_Resolution_Models}. It can be seen that in each of the high resolution models there is indeed more detail with local maxima and minima in the radial intensity profile. However, the overall trends caused by the centrall peaked enhancement of CO are preserved in each model. Therefore, the standard observations should be sufficient to detail the CO enhancement in the inner disk, despite dilution effects from the beam. 

\end{document}